\documentstyle[epsfig]{mn1}

\def\beq{\begin{equation}}
\def\eeq{\end{equation}}
\def\bey{\begin{eqnarray}}
\def\eey{\end{eqnarray}}

\def\v200{V_{200}}

\def\rh{r_{200}}

\def\et{{\it et\thinspace al.}}    %et al.%

\def\araa{{\rm ARA\&A}, }
\def\aj{{\rm AJ}, }  %Astronomical Journal%
\def\apj{{\rm ApJ}, }  %Astrophysical Journal%
\def\apjs{{\rm ApJS}, }  %Astrophysical Journal Supplements%
\def\mn{{\rm MNRAS}, }      %Monthly Notices of the Royal%
\def\aa{{\rm A\&A}, }     %Astronomy & Astrophysics%
\def\spose#1{\hbox to 0pt{#1\hss}}
\def\lta{\mathrel{\spose{\lower 3pt\hbox{$\mathchar"218$}}
     \raise 2.0pt\hbox{$\mathchar"13C$}}}
\def\gta{\mathrel{\spose{\lower 3pt\hbox{$\mathchar"218$}}
     \raise 2.0pt\hbox{$\mathchar"13E$}}}

\def\clock{\count0=\time \divide\count0 by 60
     \count1=\count0 \multiply\count1 by -60 \advance\count1 by \time
     \number\count0:\ifnum\count1<10{0\number\count1}\else\number\count1\fi}

\title[Formation of Hubble Sequence]{The Formation of the Hubble Sequence of Disk Galaxies: The Effects of Early Viscous Evolution}

\author[B. Zhang and R.F.G. Wyse]{Bing Zhang and Rosemary F.G. Wyse
\thanks{E-mail: bingz@pha.jhu.edu (BZ); wyse@pha.jhu.edu (RFGW)}  \\
Department of Physics and Astronomy, Johns Hopkins University, 3400 N.Charles Street, Baltimore, MD 21218, USA }

\date{\today}
\begin{document}
\maketitle

\begin{abstract}

We investigate a model of disk galaxies  whereby viscous evolution of 
the gaseous disk drives material inwards to form a proto-bulge.   
We start from the  standard picture of disk formation through the 
settling of gas into a dark halo potential well, with the disk initially 
coming into 
centrifugal equilibrium with detailed conservation of angular momentum. 
We derive generic analytic solutions for the disk-halo system 
after adiabatic compression of the dark halo, with 
free choice of the input virialized dark halo density profile and of the  
specific angular momentum distribution.
We derive limits on the final density profile of the halo in the central  
regions.   Subsequent viscous evolution of the disk is modelled by a 
variation of the specific angular momentum distribution of the disk, 
providing analytic solutions to the final disk structure.  The assumption 
that the viscous evolution timescale and star formation timescale are 
similar leads to predictions of the properties of the stellar components. 
Focusing on small `exponential' bulges, ones that may be formed through 
a disk instability, we investigate the relationship 
between the assumed initial conditions, such as halo `formation', or 
assembly, redshift $z_f$, 
spin parameter $\lambda$, baryonic fraction $F$, and final disk properties 
such as global star formation timescale, gas fraction, and bulge-to-disk 
ratio. We find that the present properties of disks, such as the scale 
length, are compatible with a higher 
initial formation redshift if the re-distribution by 
viscous evolution is included than if it is ignored.  We also quantify the  
dependence of final disk properties on 
the ratio $F/\lambda$, thus including the possibility that the baryonic 
fraction varies from galaxy to galaxy, as perhaps may be inferred from the 
observations.

\end{abstract}


\section{Introduction}

 The current picture of disk galaxy formation and evolution has as its 
basis the dissipative infall of baryons within a dominant dark halo 
potential well (White \& Rees 1978).  
The collapse and spin-up of the baryons, with angular momentum 
conservation, can provide an explanation for many of the observed 
properties of disks, with the standard initial conditions of baryonic 
mass fraction $F \sim 0.1$ and dark halo angular momentum parameter 
$\lambda \sim 0.07$ (Fall \& Efstathiou 
1980; Gunn 1982; Jones \& Wyse 1983; Dalcanton, Spergel \& Summers 1997; 
Hernandez \& Gilmore 1998; Mo, Mao \& White 1998; van den Bosch 1998). 
Galaxies such 
as the Milky Way which have an old stellar population  in the disk, must, 
within the context of a hierarchical-clustering scenario,  
evolve through only quiescent merging/accretion, so as to avoid excessive 
heating and disruption of the disk (Ostriker 1990).  Further, the 
merging processes with significant substructure cause 
angular momentum transport to the outer regions, 
which must somehow be suppressed to allow the formation of 
extended disks as observed (e.g. Zurek, Quinn \& Salmon 1988; Silk \& Wyse 
1993; Navarro \& Steinmetz 1997).  

Thus here we adopt the simplified picture 
that disk galaxies form from  
smooth gaseous collapse to centrifugal equilibrium, 
within a steady dark halo potential.  We discuss 
where appropriate below how this may be modified to take account of 
subsequent infall, or earlier star formation. 
Our model incorporates the adiabatic response of the dark halo to the disk 
infall, and we provide new, more general, analytic solutions for the 
density profile, given a wide range of initial density profiles and 
angular momentum distributions.  We provide new insight 
into the `disk-halo' conspiracy within the 
context of this model, demonstrating how an imperfect conspiracy is 
improved by the disk-halo interaction.  We explicitly include 
subsequent viscous evolution of the gas disk to provide the exponential 
profile of the stellar disk, and develop 
analytic expressions that illustrate the process.  The resulting radial 
inflow builds up the central regions of 
the disk and we investigate the properties of `bulges' that may form as a 
consequence of instabilities of the central disk.  We derive new 
constraints on the characteristic redshift of disk star formation. 
We obtain a simple relation connecting the initial
conditions, such as spin parameter and baryonic mass fraction, to the 
efficiency of viscous evolution and star formation.

\section{The Disk Galaxy Formation Model}                   

In this section we shall derive the mass profiles of disk and halo after the collapse of the baryons. We shall follow earlier treatments of disk galaxy formation (e.g. 
Mo, Mao \& White 1998) by assuming that the 
virialized dark halo, mixed with baryonic gas, is `formed' -- or at least 
assembled --  at redshift $z_f$. 
This virialized halo has a limiting radius $r_{200}$ within which
the mean density is $200 \rho_{crit} (z_f)$, and contains a baryonic 
mass fraction  $F$. Then 
\beq  \label{rh_h}
\rh ={V_{200} \over 10 H(z_f)};\,\,\,\,\,\,\,
M_{tot}={V_{200} ^2\rh \over G}={V_{200} ^3 \over 10 G H(z_f)} ,
%\eqno(2)
\eeq 
where $H(z_f)$ is the value of the Hubble parameter 
at redshift $z_f$, $M_{tot} (z_f)$ is 
the total mass within virialized radius $r_{200}$, and 
$V_{200}$ is the circular velocity at $r_{200}$. 

The baryonic gas cools and settles into a disk, causing the dark halo to
contract adiabatically (Blumenthal {\it et al.} 1986).  The specific 
angular momentum distribution of the gas is assumed to be 
conserved during these stages.  We shall include below the subsequent 
re-arrangement of the disk due to angular momentum transport.  This we 
investigate by variation of the disk angular momentum 
distribution function, choosing an appropriate analytic functional form.  
Let  
$m_d (r)$ and $m_h(r)$ respectively 
denote  
the fraction of the total baryonic mass, and total dark mass, that is 
contained within radius $r$, and denote the  
baryonic mass angular momentum distribution 
function by 
\beq
m_d \left( <j \right)= f \left( j/j_{max} \right) ,
\eeq
where $j_{max}$ is the maximum specific angular momentum of the disk.

We will be requiring that the functional form, $f(j/j_{max})$, vary as the 
disk evolves, and it is convenient to introduce the notation 
$\ell \equiv j/j_{max}$ and define  
\beq
c_f \equiv 1-\int_0^1 f \left( \ell \right) d \ell ,
\eeq
which represents the area above the angular momentum distribution function 
curve $f(\ell)$ for $0 \leq \ell \leq 1$.  We will mimic the effects of 
viscous evolution by decreasing the value of $c_f$ in our evolving disk 
models in section 4 below. 

In terms of this parameter the total disk angular momentum  is:
\begin{eqnarray}
J_d &=& F J_{tot}\,\,\, =\,\,\, M_d \int_0^{j_{max}} j \frac{dm_d}{dj} dj \nonumber\\
&=& M_d j_{max} 
\left( 1-\int_0^1 f \left( \ell \right) d \ell \right ) 
\,\,\, =\,\,\, F M_{tot} j_{max} c_f ,
\end{eqnarray}
with $M_d = FM_{tot}$. 
Thus $j_{max} c_f$ is the average specific angular momentum of the disk 
material.

The specific angular momentum of the disk material is assumed to 
follow that of the dark halo, but in general will not be a simple analytic 
function (e.g. Quinn \& Binney 1992). For illustration, 
we adopt an analytic monotonic increasing function $f(b,\ell)$ containing 
a free parameter $b$, with $ 0 \leq b \leq 1$. 
We require the initial angular momentum distribution to be scale free,  
representing the angular momentum distribution of the virialized halo, and will adopt $f(b=0,\ell) = \ell^n$.  
We shall vary the value of the parameter $b$ to mimic 
the effects of viscous evolution on the angular momentum distribution.

The total energy is:
\beq
E_{tot} = - \frac{\epsilon_0 G M_{tot}^2}{2 r_{200}}= - \frac{\epsilon_0 M_{tot} V_{200}^2}{2} ,
\eeq
where $\epsilon_0$ is a constant of order unity, depending on the dark halo density 
profile, and since it is constant for any specific halo model, we can 
take $\epsilon_0=1$ without loss of 
generality. The spin parameter $\lambda$ is by definition
\beq
\lambda \equiv J_{tot} | E_{tot} |^{1/2} G^{-1} M_{tot}^{-5/2}.
\eeq
Thus the mean specific angular momentum of the disk material 
may be expressed as 
\beq
c_f j_{max} =\sqrt{2} \lambda V_{200} r_{200}. 
\eeq
Assuming spherical symmetry, the  rotationally-supported disk has mass 
profile given by 
\beq
m_d = f \left( j/j_{max} \right) = 
f \left( \frac{\sqrt {G M_{tot} \left( m_d \right) r \left( m_d\right)}}{j_{max}} \right) = f \left( \ell \right) ,
\eeq
while the initial virialized halo mass profile is 
\beq
g(R_{ini})  =  m_h(R_{ini}) = M_{ini}(R_{ini}) /M_{tot} ,  
\eeq
with $R \equiv r/r_{200}$.

\subsection{Constraints on the Final Dark Halo Profile and Mass Angular 
Momentum Distribution Function}

The above equations describe the disk and halo just upon the settling of the 
gas disk to the mid-plane, prior to the 
subsequent adiabatic compression of the halo. 
A self-consistent calculation of the modified disk and halo density profiles 
may be made by consideration of 
the adiabatic 
invariance of the angular action,  
$I_\theta \equiv \int v_\theta \cdot r d \theta  = \sqrt{G M_{tot}(r) r}$, 
together with the 
assumption of no shell crossing (cf. Blumenthal {\it et al.} 1986). 

Suppose a dark matter particle initially at $r_{ini}$ 
finally settles at $r(m_d)$, 
the radius within which the 
dark halo mass fraction is $m_h$. Then under adiabatic invariance the disk 
mass profile, $m_d$, and the halo mass profile, $m_h$, are related by: 
\begin{eqnarray}
G M_{tot} \left( m_d \right) r \left( m_d \right) 
& =&  G M_{tot} \left( m_h \right) r \left( m_h \right) \\
& =& G M_{ini} \left( m_h \right) r_{ini} \left( m_h \right) \\
& =& G M_{tot} m_h r_{200} g^{-1} \left( m_h \right) ,
\end{eqnarray}
where $g^{-1} ( m_h) $ is the inverse function of $g(R)$, the initial 
virialized 
halo mass profile.

Further manipulation of these relations is simplified by introduction of the 
parameter $\xi$, given by 
\beq
\xi \equiv  \frac{\sqrt{G M_{tot} r_{200}}}{j_{max}} = \frac{c_f}{\sqrt{2} 
\lambda}. 
\eeq
Generally $\xi$ is a quantity that is closely related to the overall disk 
collapse factor.

From equations (8) and (12), we have 
\begin{eqnarray}
m_d &=& f(\ell) ,\\
\ell   &=&   \xi  \left( m_h g^{-1} ( m_h ) \right)^{1/2}, 
\end{eqnarray}
where $\ell$ is the normalized specific angular momentum. Again, $ 0
\leq \ell \leq 1$, and $\ell=1$ corresponds to the maximum specific
angular momentum of the disk, which occurs at the edge of the disk,
equivalently at the disk cutoff radius.  The fraction of the dark
matter contained within the disk thus has a maximum value, $m_{hc}$,
given by $\ell = 1$ in the above equation, and for radii with $m_h \geq
m_{hc}$, $m_d =1$.

To illustrate the physical meaning of these parameters, consider the 
rigid singular isothermal halo, for which $g(m_{hc}) = m_{hc} = R_c$.  Then 
from equation (15),  $ R_c =  R(\ell=1) = 1/\xi$ which 
corresponds to the disk cutoff radius. 
Thus in this case, remembering that $R$ is the normalized radius, 
$\xi =1/R_c$ is the disk collapse factor.

Up to now we know the mass profile of disk and halo after collapse, in
terms of the normalized specific angular momentum $\ell$, as given in
equations (14) and (15), for given forms of the angular momentum
distribution function, $f$, and initial virialized dark halo mass profile,
$g$. Next we shall obtain the relation between $\ell$
and radius $R$, to complete the derivation of the mass profiles of disk and
halo after collapse.

Returning to a general halo density profile, 
the total mass contained within the radius corresponding to $m_h$ is
\beq
M (m_h) = M_{tot} \left( (1-F) m_h + F m_d \right). 
\eeq
From equations (9) - (12) and (16), we have
\begin{eqnarray}
R & = &  \frac{G M_{ini} (m_h) r_{ini} (m_h)}{G M(m_h) r_{200}} \\
  & = & \frac{m_h g^{-1} (m_h)}{(1-F) m_h + F m_d}.
\end{eqnarray}
Introducing the radius variable $x$ and coefficient $c_0$ as
\begin{eqnarray}
x & \equiv & \xi (1-F) R, \\
c_0 & \equiv &\xi F/(1-F),
\end{eqnarray}
we may finally derive the 
functional dependences on $\ell$ of the disk mass $m_d$, of 
the 
halo mass $m_h$, and of the  
radius $x$:
\begin{eqnarray}
m_d &=& f(\ell), \\
m_h g^{-1}(m_h) &=& \frac{\ell^2}{\xi^2}, \\ 
x  &=& \frac{\ell^2}{\xi m_h (\ell) + c_0 f(\ell)}.
\end{eqnarray}
Thus $\ell$ can be thought of as a normalized radius. 
As we shall see later, $c_0$ is a measure of the compactness of the 
final collapsed disk 
due to the competition between the spin parameter $\lambda$ and 
the baryonic mass fraction $F$.

These equations (21) - (23) can be used to derive disk and halo properties 
for a free choice of virialized halo profile $g(R)$ and 
angular momentum distribution function $f(\ell)$.
Within the disk cutoff radius, with  $ 0 \leq \ell \leq 1$, 
the disk surface density, circular velocity and the disk-to-dark mass 
ratio as function of radius $\ell$ or $R$ have the generic forms:
\begin{eqnarray}
\Sigma_d &=& \frac{10 H(z)F V_{200}}{2 \pi G} 
\frac{1}{R} \frac{df}{d\ell} \frac{d \ell}{dR}, \\
V_c &=& \frac{V_{200} \ell} {\xi R}, \\
\frac{M_d(\ell)}{M_h(\ell)} &=& \frac{c_0 f(\ell)} {\xi m_h(\ell)},
\end{eqnarray}
where $M_d(\ell) = M_d(\ell=1)m_d(\ell) = M_d m_d(\ell)$ and $M_h(\ell)$ is 
defined similarly. 

The circular velocity at radii beyond the disk cutoff, but within the halo, is given by:
\begin{eqnarray}
V_c  &=& \frac{V_{200} \sqrt{m_h g^{-1}(m_h)}}{R}, \\
R &=& \frac{m_h g^{-1}(m_h)}{(1-F) m_h + F}.
\end{eqnarray}

\begin{figure}
\centerline{\psfig{file=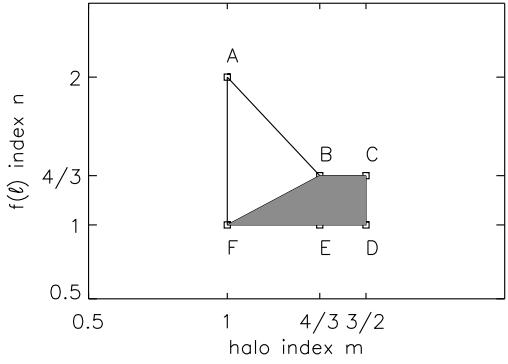,width=2.6in,angle=0}}
\hspace{0.5cm}
\caption{The power-law approximations for the initial virialized 
halo mass profile and angular momentum distributions are
$m_h(\ell) \sim \ell^m$ and  $f(\ell) \sim \ell^n$
for small $\ell$.  
The shaded region is the allowed parameter space for these models, 
constrained by the surface-density profile, rotation curve and disk-to-halo 
central mass ratio. The line ABC is the maximum angular-momentum index 
consistent with a disk surface density profile in the central regions 
that declines with increasing radius, while the line DEF is the  
minimum angular-momentum index consistent with a finite value of the 
central circular velocity.  The line BF denotes the maximum values of $n$ 
consistent with a non-zero central disk-to-halo mass ratio.  
The value $m=1$ corresponds to the singular isothermal sphere, the value $m=4/3$ 
corresponds to the Hernquist (1990) and to the Navarro, Frenk \& White 
(1997) profiles, while  the value $m=3/2$ corresponds to the 
non-singular isothermal sphere with  a constant-density core.}
\end{figure}

Armed with these relations, one may now look at various 
initial virialized halo density profiles and angular momentum distributions, 
and determine the allowed parameter space from observed properties of disk 
galaxies.  It is convenient to adopt power-law approximations for 
the initial virialized 
halo mass profile and angular momentum distributions, such that   
$m_h(\ell) \sim \ell^m$ and  $f(\ell) \sim \ell^n$
for 
small $\ell$.  Figure 1 shows the location of various fiducial models in the 
plane of these power law indices $m$ and $n$;  the value $m=1$ corresponds 
to the singular isothermal sphere,  the value $m=4/3$ 
corresponds to the Hernquist (1990) and to the Navarro, Frenk \& White 
(1997) 
profiles, while  the value $m=3/2$ corresponds to the 
non-singular isothermal sphere with  a constant-density core.  These 
profiles span the range of dark-halo profiles suggested by theory, and 
plausibly consistent with observations. 
The shaded region is the allowed parameter space for these models, 
constrained by the surface-density profile, rotation curve and disk-to-halo 
central mass ratio. 
The line ABC is the maximum angular-momentum index consistent with a disk 
surface density profile in the central regions 
that declines with increasing radius,  
while the line DEF is the  
minimum angular-momentum index consistent with a finite value of the central 
circular velocity.  The line BF denotes the maximum values of $n$ consistent 
with a non-zero central 
disk-to-halo mass ratio.  

This power-law approximation has an initial 
virialized halo density profile at small radius as 
$\rho_{h,ini}(R) \sim R^{\frac{m}{2-m}-3}$ (seen by solution of (22) for the form 
of $g$). Solving equations (21) - (23) 
above gives the corresponding profile after adiabatic infall.
 In the central region, where the disk dominates the gravitational potential 
(i.e. $c_0 f(\ell) \gg \xi m_h(\ell)$),   
the halo density profile is $\rho_h(R) \sim R^{\frac{m}{2-n}-3}$. 
Note that in the region where the dark halo dominates the gravitational 
potential (i.e. $c_0 f(\ell) \ll \xi m_h(\ell)$), 
the halo density profile is essentially unaffected by the disk, as expected. 

For the case $n=m$, the central halo density profile is unaffected 
since the disk mass density 
profile and the initial virialized 
halo density profile have the same dependence on $\ell$. 
The viable models within the shaded region have $n \leq m$, so that 
the final halo density profile in the central, disk-dominated region 
should be steeper than its initial virialized profile 
in this region, not surprisingly. 
Thus the final halo profile for 
these models ranges from $\rho_h \sim R^{-0.75} $ 
to $\rho_h \sim R^{-2}$. It is interesting to note 
that $\rho_h \sim R^{-0.75} $, the outcome of  an initial
virialized halo with a constant density core  responding 
to the settling of a disk with angular momentum index $n = 4/3$, 
corresponds to the de-projected 
de~Vaucouleurs central density profile.  Thus provided the virialized halo 
does not have a declining density profile with decreasing radius,  which is 
unphysical, the final dark halo 
cannot have a constant density core, at least in the very central region 
where the disk dominates, but should be cuspy.  

\subsection{The Singular Isothermal Sphere}

The singular isothermal sphere provides a virialized halo density profile 
that is the  most tractable analytically, and we can obtain some important 
scaling relations without having to specify the angular momentum 
distribution $f(\ell)$; aspects of the analysis of 
this profile should hold in general, and provide insight. 

The final disk and halo mass profiles are given by solution of:
\begin{eqnarray}
x &=& \frac{\ell}{1 + c_0 \frac{f(\ell)}{\ell}}, \\
m_h &=&  \ell/\xi, \\
m_d &=& f(\ell),
\end{eqnarray}
where $ 0 \leq \ell \leq 1$ and $c_0$ and $\xi$, the parameters describing the compactness of the collapsed disk and its collapse factor, are defined above in 
equations (20) and (13). 

The collapse factor, defined as the ratio of the pre-collapse radius
$r_{200}$ to the cutoff disk radius $r_c$ at $\ell =1$ (to be
distinguished from the final disk scale length) is \beq
\frac{r_{200}}{r_{c}} = \xi (1-F)(1+c_0)= \frac {c_f}{\sqrt{2}
\lambda} \left( 1-F + \frac{c_f}{\sqrt{2}} \frac{F}{\lambda} \right),
\eeq where $c_f$, defined in equation (3), is a measure of the shape
of the angular momentum distribution, and small values of $c_f$ mean
steeply-rising angular momentum distributions.

\begin{figure}
\centerline{\psfig{file=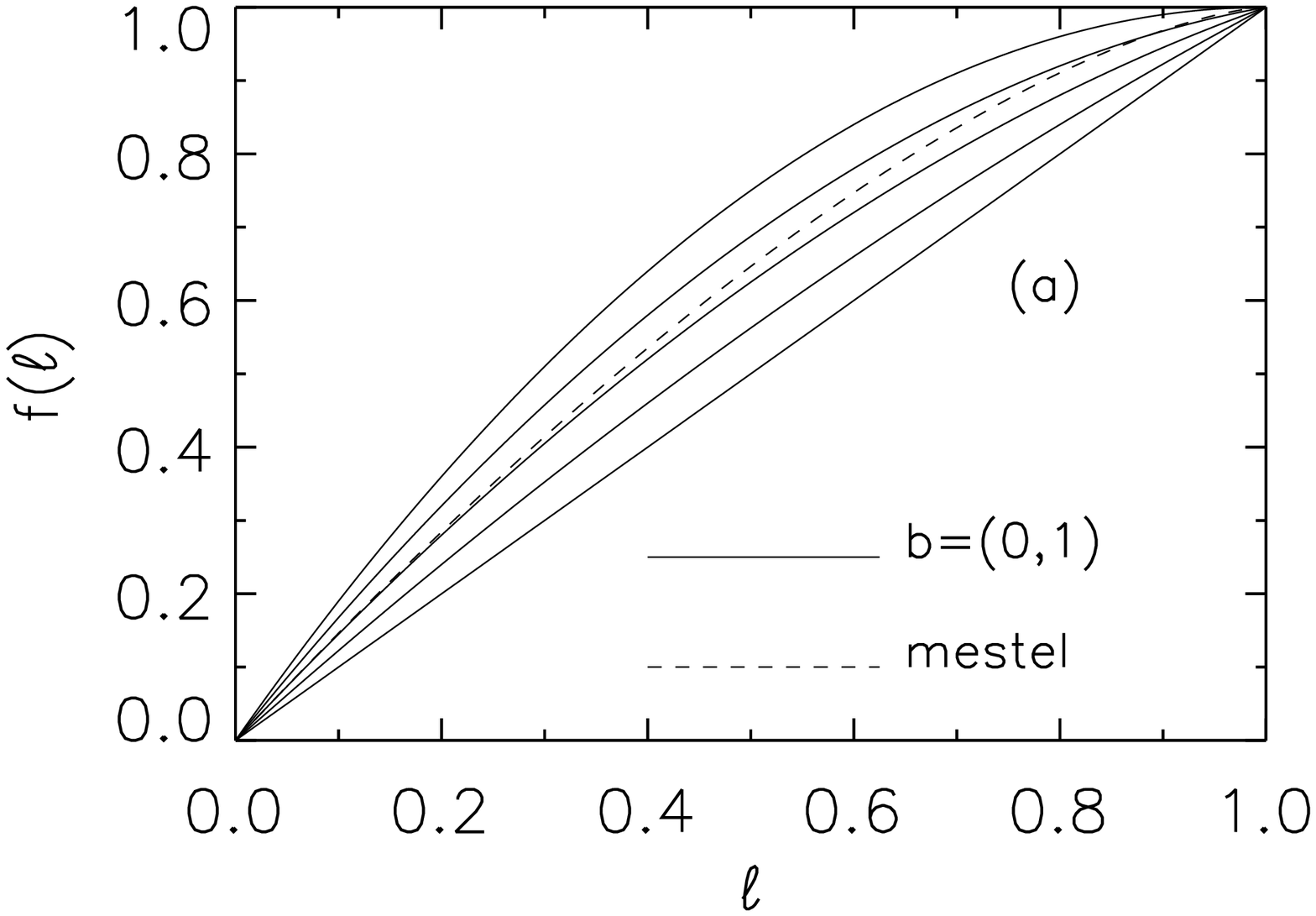,width=2.6in,angle=0}}
\vspace{0.5cm}

\centerline{\psfig{file=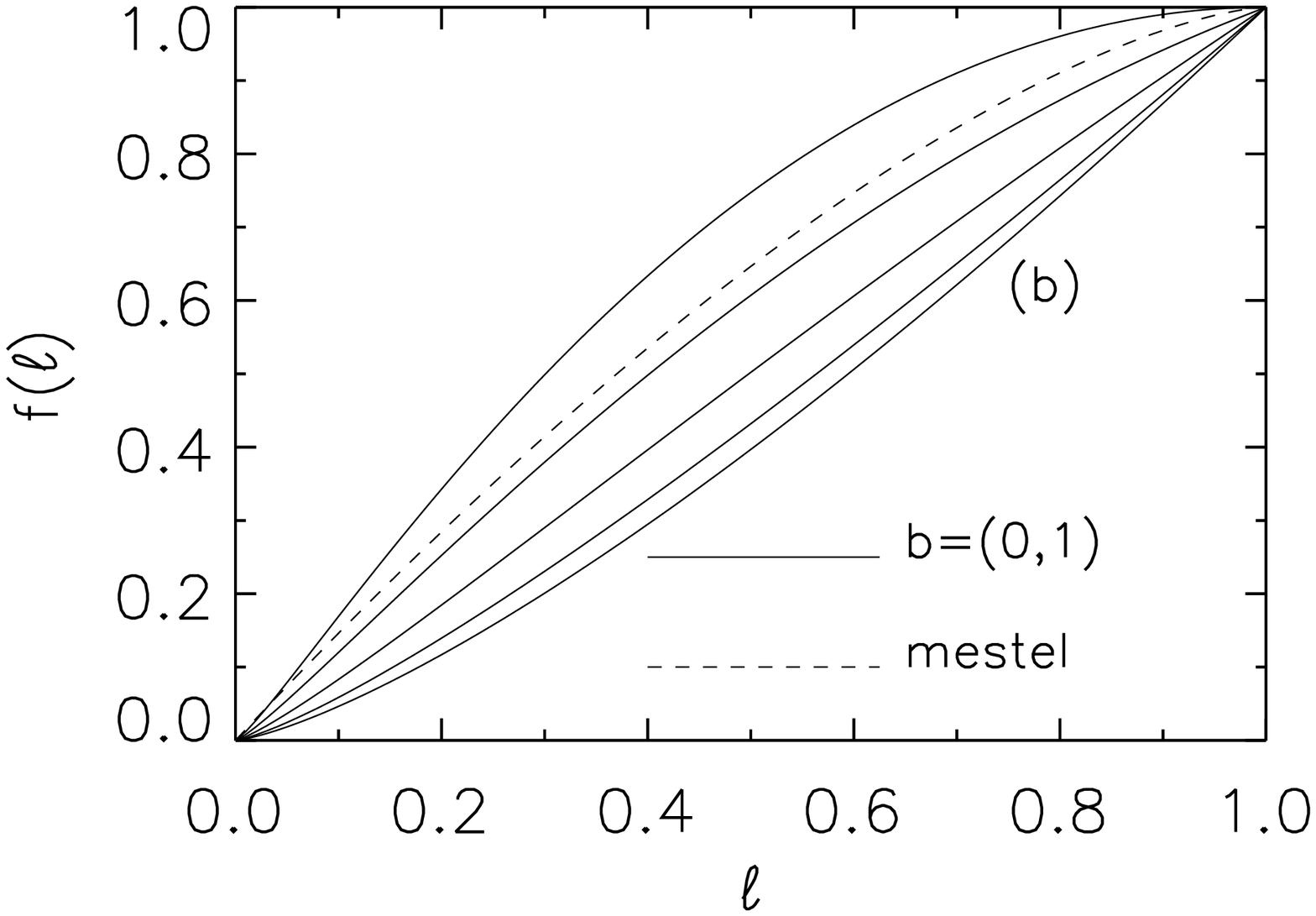,width=2.6in,angle=0}}
\vspace{0.5cm}
\caption{The angular momentum distribution function for the Mestel profile, 
$f(\ell)=1-(1-\ell)^{3/2}$, is shown by the dashed line. 
(a) The angular momentum distribution function is
$f(b,\ell)= (1+b) \ell -b \ell^2 $ for models on line FED in Figure 1. 
The different curves correspond to $b= 0, 1/4, 1/2, 3/4, 1$.
(b) The angular momentum distribution function is
$f(b,\ell)=(1+10b)\ell^{4/3}-10b\ell^{22/15}$  
for models on line BC in Figure 1.  The different curves correspond to $b= 0, 0.1, 0.3, 0.6, 1$.}
\end{figure}

Typical angular momentum distributions for disks are shown in Figure 2 (a,b) 
and discussed below. Here one should just note that a value  $c_f \sim 0.5$ 
is reasonable.  With this, and 
$\lambda = 0.06$ and $F = 0.1$, we obtain $\xi = 5.9 $ 
and 
$c_0 = 0.66$. The collapse factor as defined here is then about a factor of 
9.  For fixed angular momentum distribution $c_f$, the collapse
factor depends not only on $1/\lambda$ but also on 
compactness 
$c_0 \propto F/\lambda$. This is consistent with the results of 
previous studies
of the collapse factor in two extreme cases (Jones \& Wyse 1983; Peebles 1993): 
if the final 
disk is so self-gravitating that the value of $F/\lambda$ corresponds to 
$c_0 \gg 1$, then 
the collapse factor is $\propto F/\lambda^2$;  on the other hand 
if the final disk is sufficiently far from self-gravitating, with small 
$F/\lambda$ and $c_0 \ll 1$, the collapse factor is $\propto 1/\lambda$.  
The above collapse factor relationship is valid over the entire range of 
values for the parameters $\lambda$ and $F$ (and is the most general 
relation derived to date). 

It should be noted that varying the normalized angular momentum distribution 
by varying $c_f $ changes the derived collapse factor; this is investigated 
further below.  

Comparison between the sizes of observed disks and those predicted from such 
collapse calculations provides a constraint on the redshift at which the 
collapse happened (e.g. Mo, Mao \& White 1998).  Our 
disk cutoff radius for non-trivial $f(\ell)$ may be expressed in terms of 
the initial conditions as :
\beq
r_c= \frac{\sqrt{2} \lambda V_{200}}
{10 H(z_f) c_f (1-F + \frac{c_f}{\sqrt{2}} \frac{F}{\lambda})},
\eeq
where $z_f$ is the `formation' or assembly redshift, at which 
the halo is identified to 
have given mass and circular velocity, with no mass infall after this epoch. 
One can see from this relation that 
both $\lambda$ and $F/\lambda$ are equally important; previous 
determinations considered the baryonic mass fraction fixed (Dalcanton, 
Spergel \& Summers 1997; Mo, Mao \& White 
1998).  

The 
explicit inclusion of the parameter $c_f$ allows us to take account of disk evolution, as gas is transformed to stars. 
 Adopting a
disk cutoff radius at three disk scale-lengths (e.g. van der Kruit
1987), and choosing specific values of the present-day 
stellar disk scale-length $r_d =
3.5$kpc, $\lambda =0.06$, $V_{200}=200$  kms$^{-1}$ and an
Einstein-de-Sitter Universe with Hubble constant $ 0.5< h <1$, the
above relation gives the formation redshift $1.6 < z_f < 3.1$ for
$F=0.05$ and $c_f=1/3$; $1.4 < z_f < 2.8$ for $F=0.1$ and $c_f =1/3$;
$0.9 < z_f < 2.0$ for $F=0.05$ and $c_f =1/2$; $0.7 < z_f < 1.7$ for
$F=0.1$ and $c_f=1/2$.

Consistent with previous calculations, for fixed formation redshift
smaller values of $\lambda$ can lead to smaller disk size; we have
here explicitly demonstrated that larger $F/\lambda$ can also lead to
this result. Lower values of $c_f$ also lead to higher redshift of
formation.  Thus for no viscous evolution i.e. there is no angular
momentum redistribution during the evolution of the galactic disk and
$c_f$ has a time-independent value, the formation redshift $z_f$
determined by assuming fixed initial $\lambda$ and $F$ and fixed disk
size $r_c$ is smaller than would be determined if $c_f$ could be
decreased (to mimic say viscous evolution).  Lower values of $c_f$ for
given total angular momentum content imply a larger disk scale-length;
the effect of viscous evolution is to re-arrange the disk material so
as to increase the disk scale-length. Typical values of viscosity
parameters lead to a factor of 1.5 increase in disk scale-length in a
Hubble time.  As we demonstrate below, a smaller value for $c_f$
corresponds to larger bulge-to-disk ratio. This result is consistent
with the results of Mo, Mao \& White (1998): halo and disk formation
redshift can be pushed to higher value when a bulge is included.
These trends are general, and not tied to the specific model of the
halo density profile.

One should bear in mind that the old stars in the local thin disk of the 
Milky Way have ages of at least 10~Gyr, and may be as old as the oldest 
stars in the Galaxy (Edvardsson {\it et al.} 1993); the age distribution of 
stars at other locations of the Galactic disk is very poorly-determined, but 
it is clear that a non-trivial component of the thin disk was in place at 
early times (at redshift $z > 2$ for the cosmologies considered above). 
A common assumption in previous work is that the 
mass angular momentum distribution of the disk is that of 
a solid-body, rotating 
uniform density sphere, $f(\ell) = 1-(1-\ell)^{3/2}$ (Mestel 1963). 
For this distribution, $c_f = 0.4$, and one derives a low redshift of 
formation for a galaxy like the Milky Way (Mo, Mao \& White 1998), which has 
difficulties with the observations.

\subsubsection{The Singular Isothermal Halo with Simple $f(\ell)$}

Analytic solutions to equations (29) - (31) can be obtained by assuming
a simple monotonic increasing function $f(b,\ell)$ containing one parameter $b$ 
with $ 0 \leq b \leq 1$. In order to avoid the situation where one obtains 
a trivial 
collapse factor due to a  very small amount of disk material 
at very large radius, we 
restrict the shape of the mass angular momentum distribution function $f(\ell)$ 
to avoid too shallow an asymptotic slope as $f$ approaching 1 when $\ell$ 
increases (see Figure 2). 
 From Figure 1, the singular isothermal halo at point F requires $f(\ell) \sim 
\ell$ for 
$\ell \ll 1$. A simple form consistent with this is $f(\ell) = (1+b) 
\ell -b \ell^2 $ with 
$0 \leq \ell \leq 1$ and $0 \leq b \leq 1$. The angular momentum parameter  
$c_f = (3-b)/6$. Figure 2a 
shows this angular momentum distribution function with
different values of the parameter $b$, compared with the 
mass angular momentum distribution of a solid-body, rotating 
uniform density sphere, $f(\ell) = 1-(1-\ell)^{3/2}$ (Mestel 1963). 
 The normalized total angular momentum $c_f$ 
corresponds to the area above each curve. 

Within the disk, where 
$ 0 \leq \ell \leq 1$,  
the galactic disk surface density, circular velocity and the disk-to-dark mass 
ratio as a function of radius are:
\begin{eqnarray}
\Sigma_d(\ell) = \frac{10 V_{200} H(z) F(1-F)^2\xi^2}{\pi G}  \nonumber\\
\times \frac{(1+b-2b\ell)(1+c_0+c_0b-c_0b\ell)^3}{2\ell(1+c_0+c_0b)} , 
\end{eqnarray}
\beq
V_c(\ell) = V_{200}(1-F)[1+c_0(1+b-b\ell)] ,
\eeq
\beq
\frac{M_d(\ell)}{M_h(\ell)} = \frac{(1+b-b\ell)(3-b)F}{6\sqrt{2}\lambda(1-F)}, \,\\
\eeq
where
\beq
\ell = \frac{[1+c_0(1+b)]\xi (1-F) R}{1+bc_0 \xi (1-F) R}.
\eeq
From equation (27)-(28), the circular velocity at radii beyond the disk cutoff, but within the halo, is:
\beq
V_c(R) = \frac{V_{200} (1-F)}{2} 
\left[ 1+ \sqrt{1 + \frac{4 F}{(1-F)^2 R}} \right].
\eeq

These results are plotted  in Figure 3 (a,b,c) for $\lambda =0.06$,  
$F=0.1$, $ 0 \leq \ell \leq 1$. 
The different curves correspond to $b= 0, 1/4, 1/2, 3/4, 1$.  Larger values 
of $b$ yield larger disk cut-off radii; note that the circular velocities 
for points beyond the cut-off radius of a given model may be obtained by 
forming the envelope of the values for the cut-off radius for larger values 
of $b$. 
Varying the value of the parameter $b$ changes the angular momentum 
distribution 
function similar to the effects of viscous evolution; 
the ratio of disk mass to dark halo mass increases at small radius 
with increasing $b$, which can be interpreted as due to radial inflow of 
disk material.

\subsubsection{ Halo Density Profile and Angular Momentum Combinations}

For initial virialized halo profiles other than the singular isothermal 
sphere, the collapse factor has a more general 
form: 
\beq
\frac{r_{200}}{r_{c}} = \xi (1-F)(\xi m_{hc} + c_0),
\eeq
where $m_{hc}$ is the solution of $1= \xi^2 m_{hc} g^{-1}(m_{hc})$ and is 
the mass fraction of the dark halo that is contained within the cut-off 
radius of the disk. 
The trend of the dependence of the collapse factor on the values of 
$\lambda$ and $F$ remains  the same as found above for the  
singular isothermal halo.

\begin{figure}
\centerline{\psfig{file=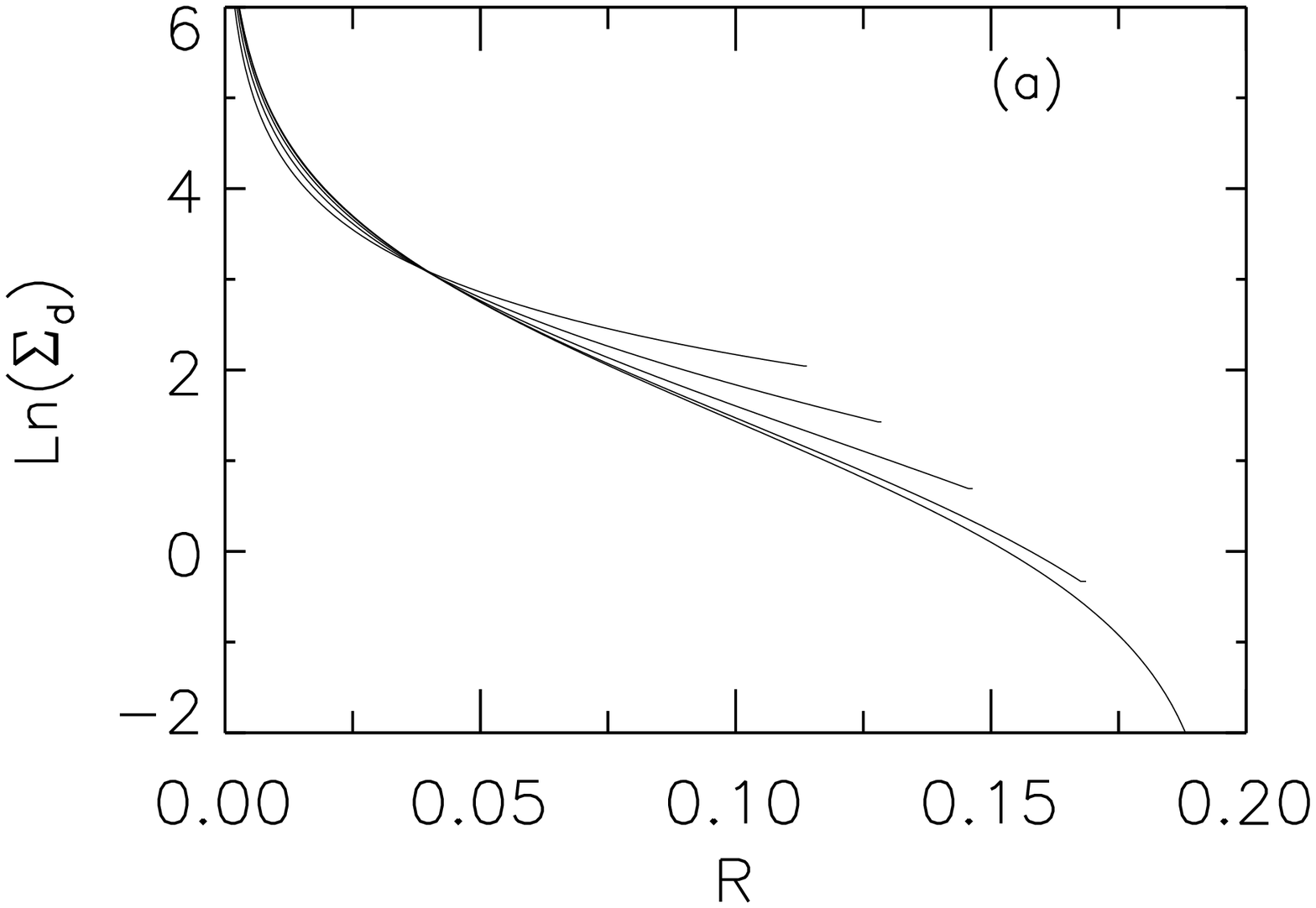,width=2.6in,angle=0}}
\hspace{0.5cm}

\centerline{\psfig{file=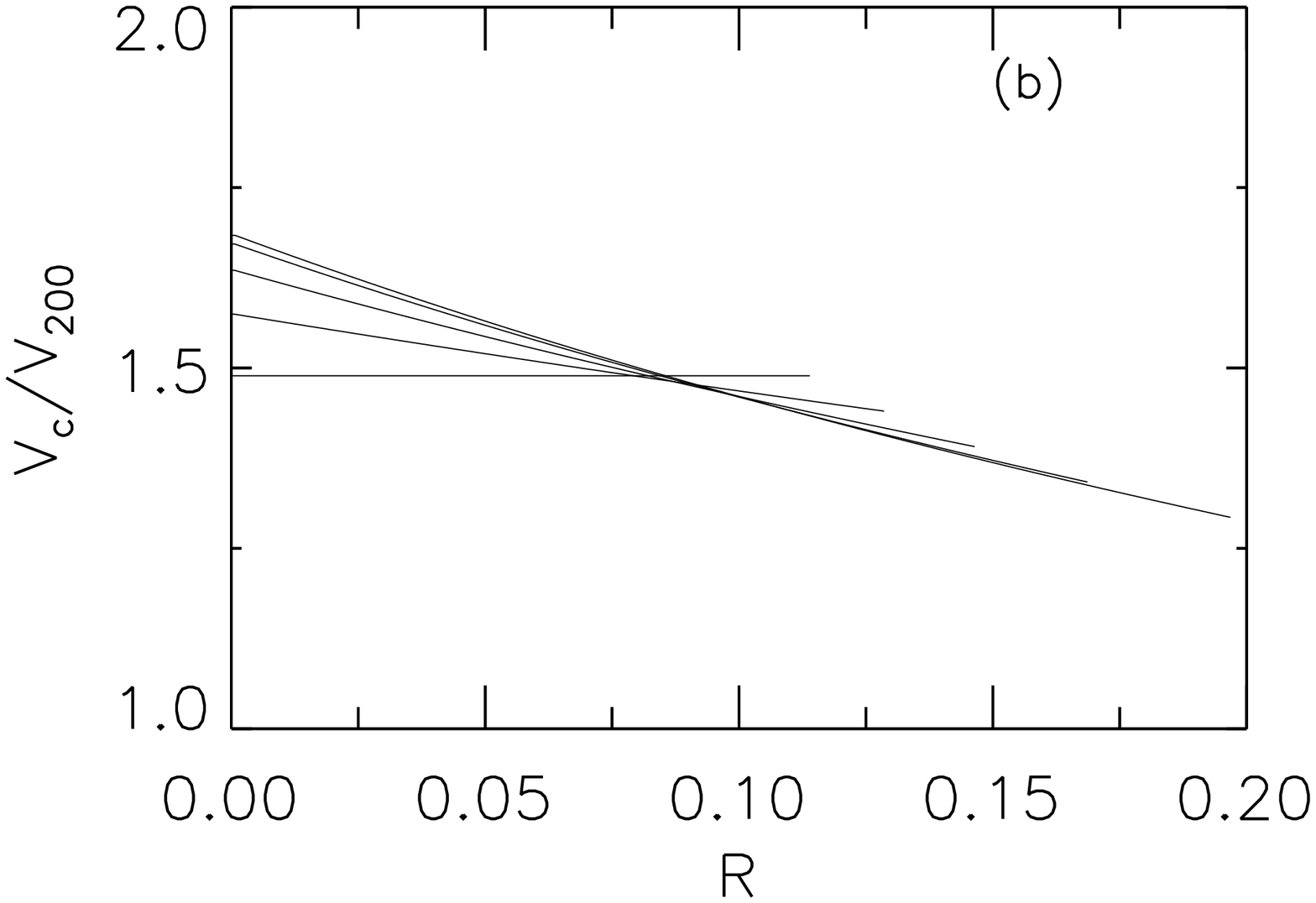,width=2.6in,angle=0}}
\hspace{0.5cm}

\centerline{\psfig{file=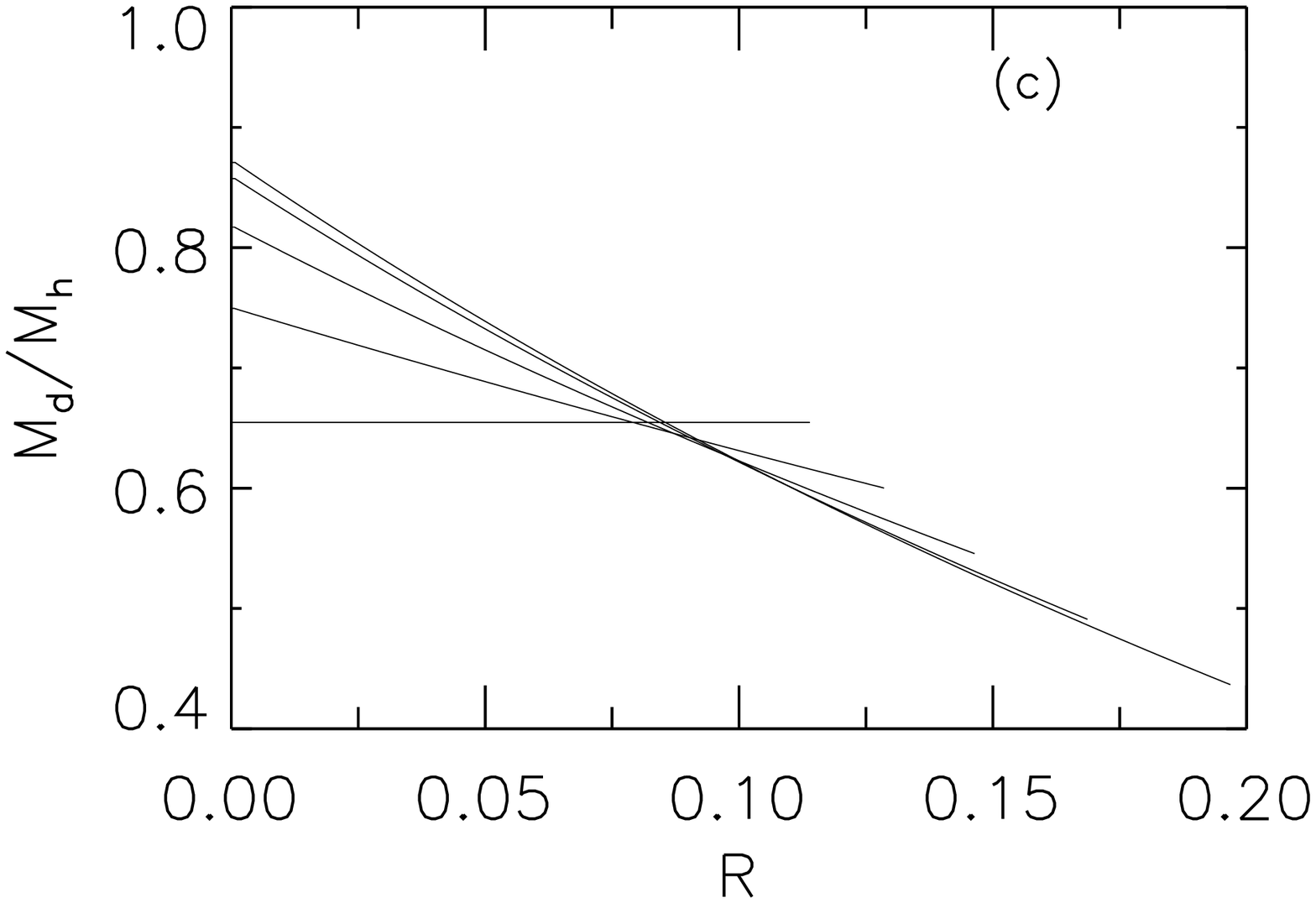,width=2.6in,angle=0}}
\hspace{0.5cm}
\caption{The model corresponding to point F in Figure 1. (a) the surface density, (b) circular velocity 
and (c) disk-to-halo mass ratio with $\lambda =0.06$ and $F=0.1$. 
The virialized halo is a singular isothermal sphere. 
The angular momentum distribution function is $f(b,\ell)= (1+b) \ell -b \ell^2 $.
The different curves correspond to $b= 0, 1/4, 1/2, 3/4, 1$. Larger values 
of $b$ yield larger disk cut-off radii.  The disk-to-halo mass ratio increases slightly when approaching the centre.}
\end{figure}

\begin{figure}
\centerline{\psfig{file=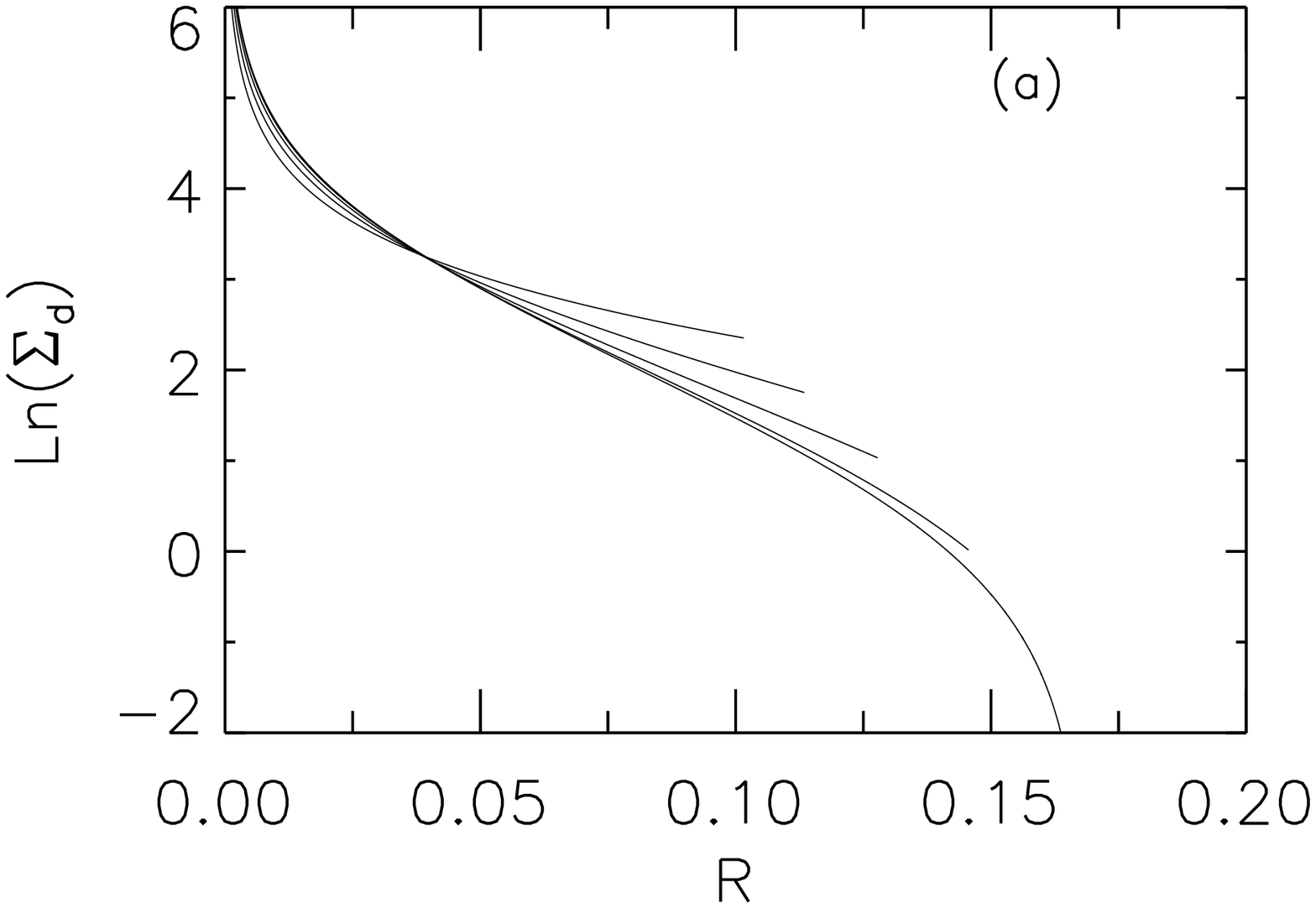,width=2.6in,angle=0}}
\hspace{0.5cm}

\centerline{\psfig{file=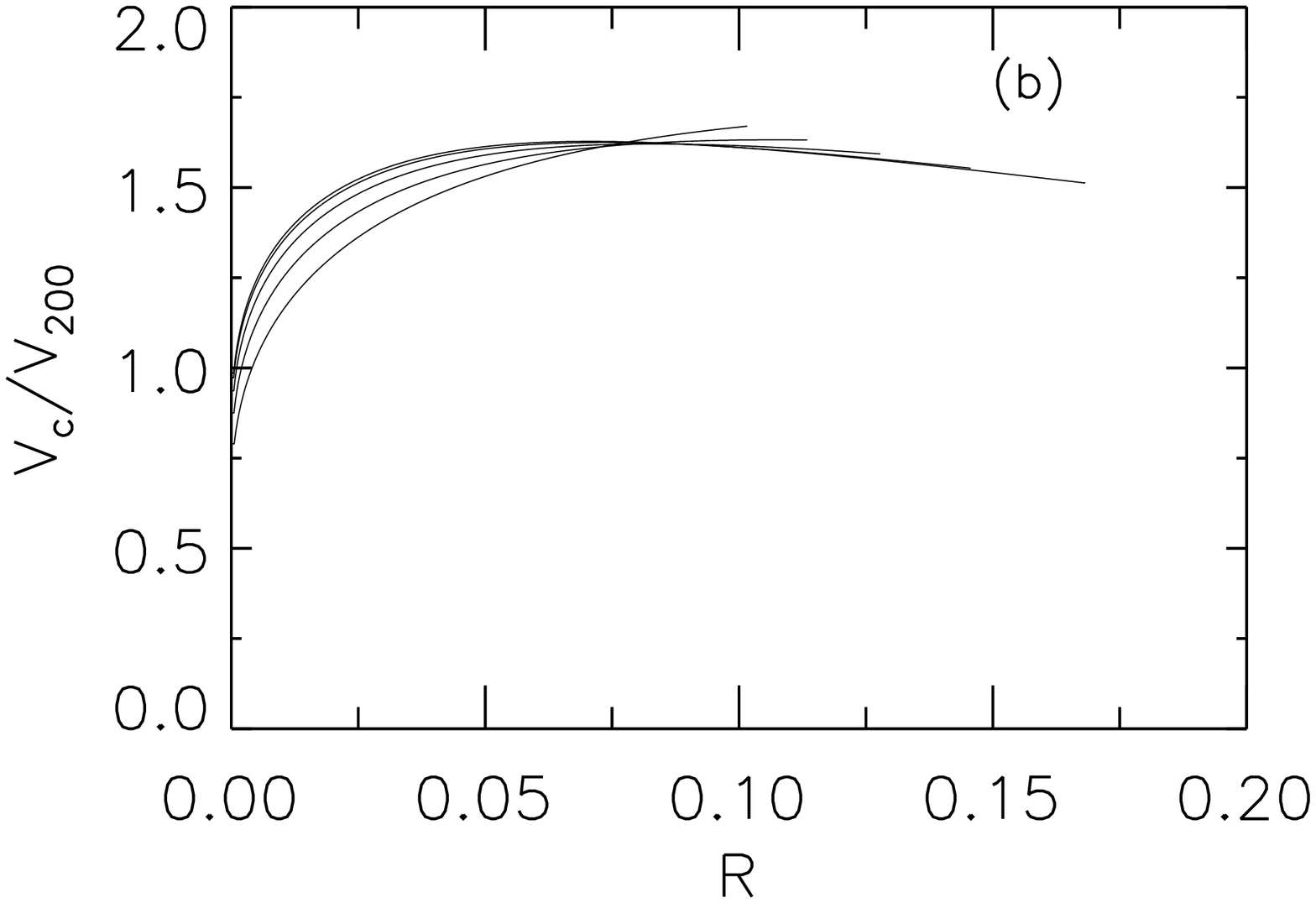,width=2.6in,angle=0}}
\hspace{0.5cm}

\centerline{\psfig{file=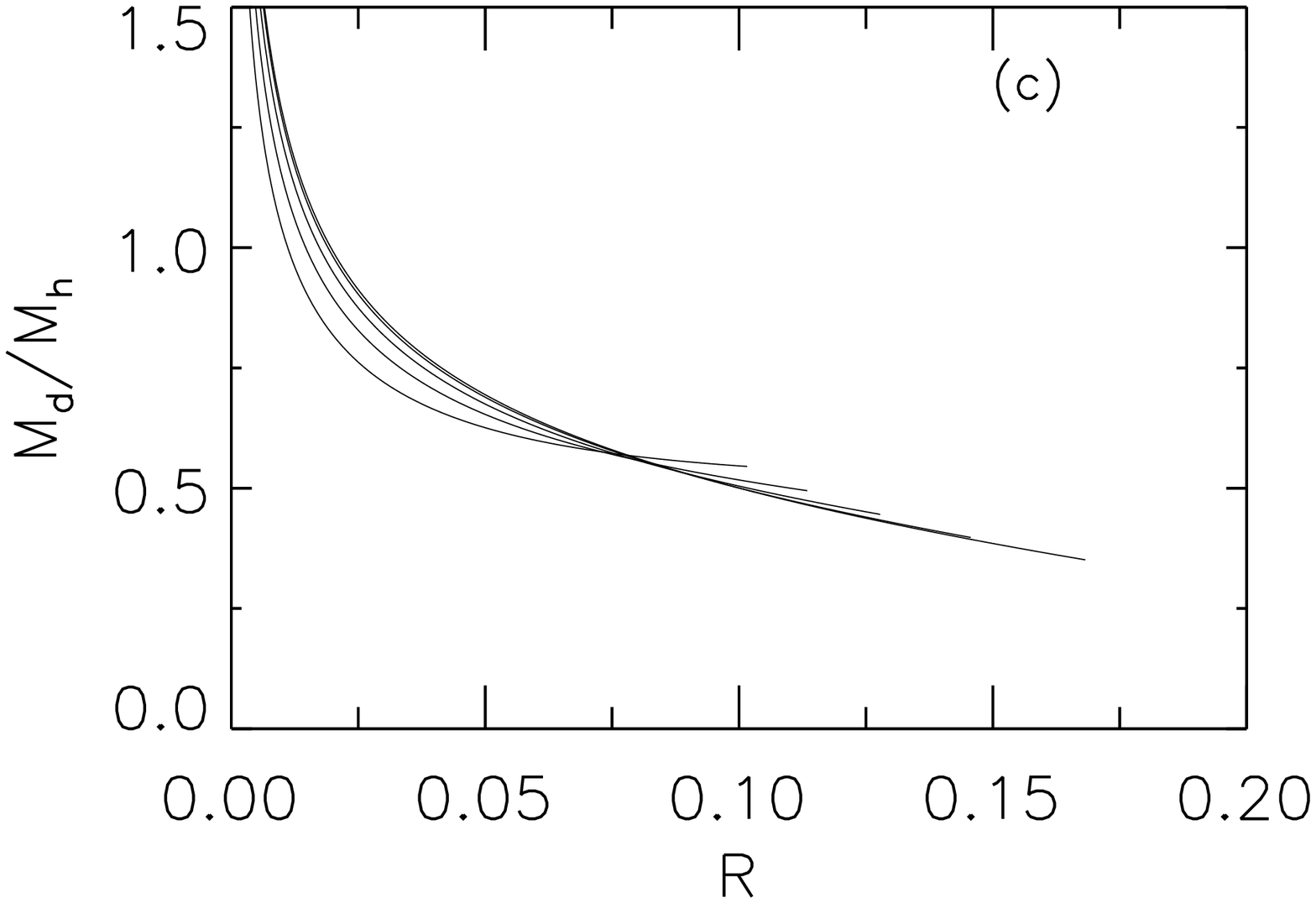,width=2.6in,angle=0}}
\hspace{0.5cm}
\caption{The model corresponding to point E in Figure 1. (a) the surface 
density, (b) circular 
velocity and (c) disk-to-halo mass ratio with $\lambda =0.06$ and $F=0.1$. 
The virialized halo is Hernquist halo with core size $c=4$. 
The angular momentum distribution function is 
$f(b,\ell)= (1+b) \ell -b \ell^2 $. 
The different curves correspond to $b= 0, 1/4, 1/2, 3/4, 1$.  
The disk-to-halo mass ratio increases significantly when approaching 
the centre.}
\end{figure}

\begin{figure}
\centerline{\psfig{file=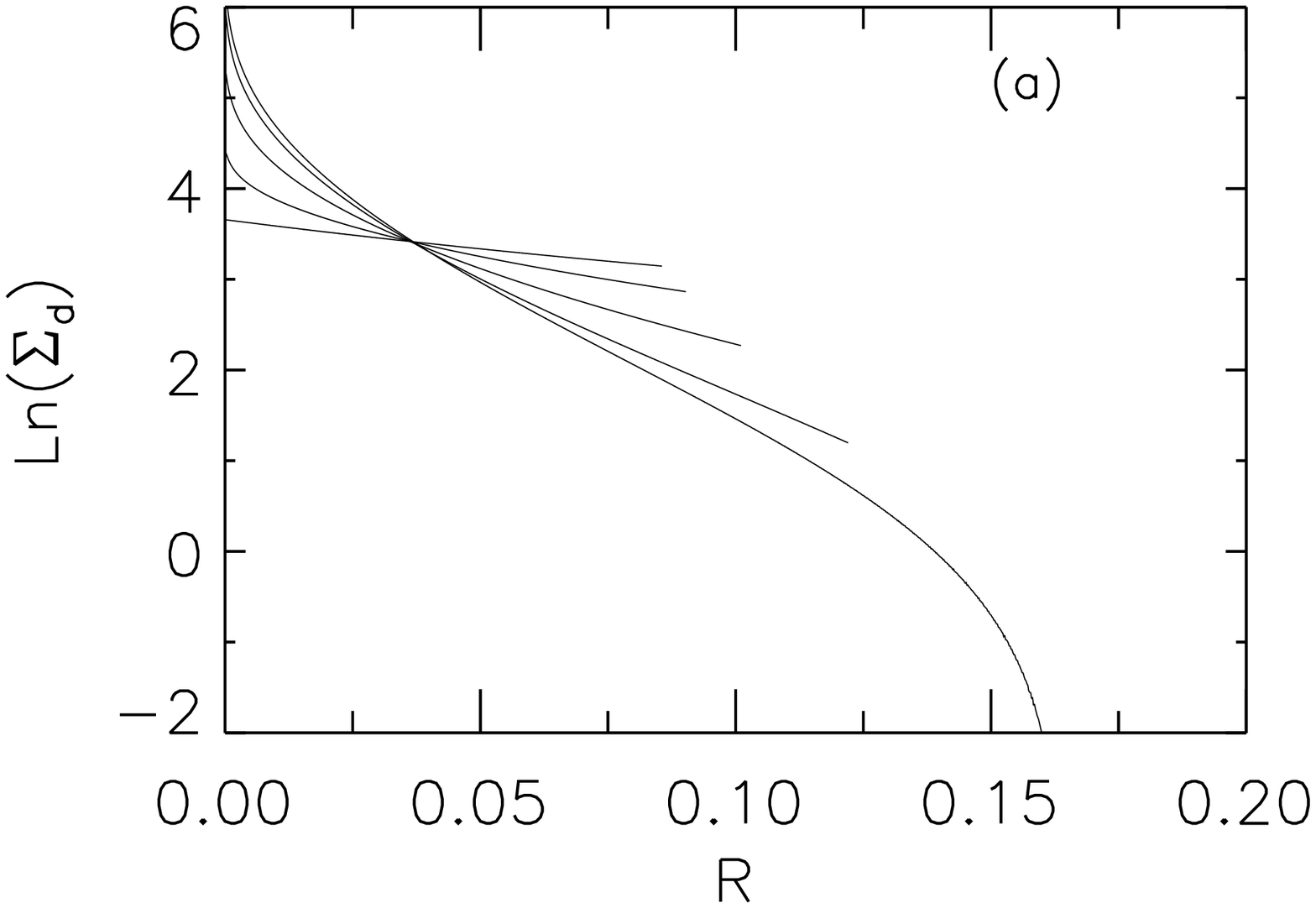,width=2.6in,angle=0}}
\hspace{0.5cm}

\centerline{\psfig{file=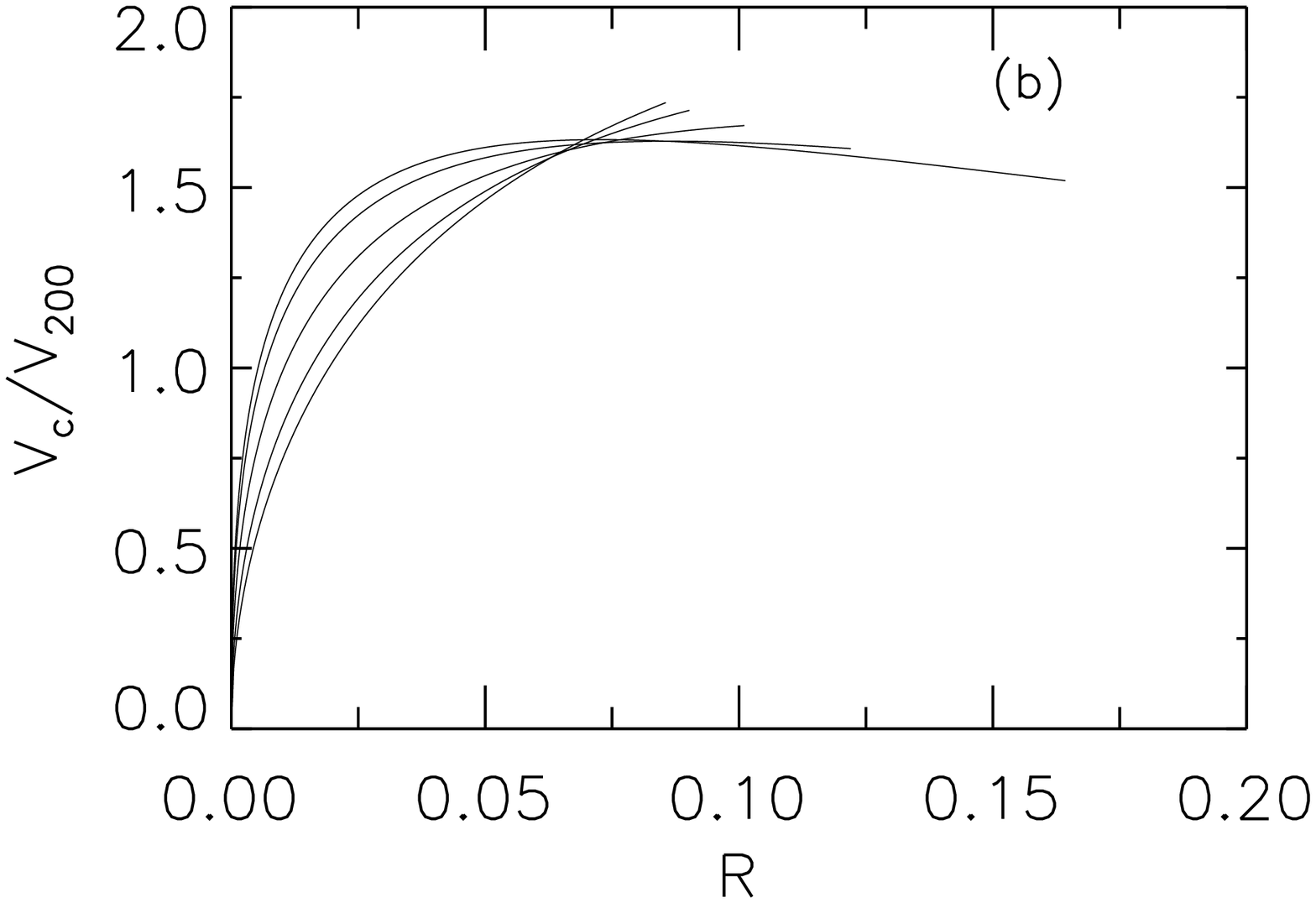,width=2.6in,angle=0}}
\hspace{0.5cm}

\centerline{\psfig{file=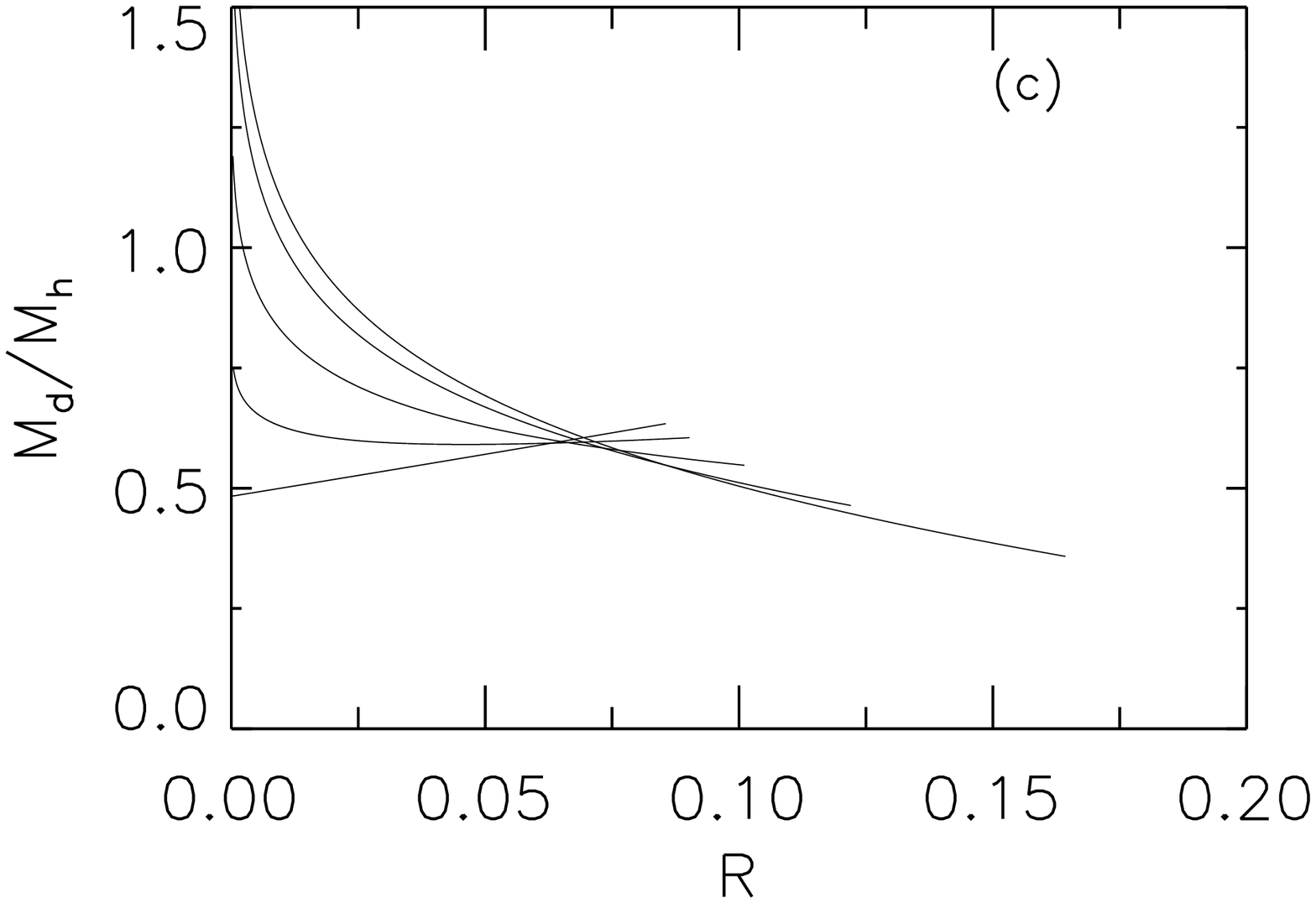,width=2.6in,angle=0}}
\hspace{0.5cm}
\caption{The model corresponding to point B in Figure 1. 
(a) the surface density, (b) circular velocity and 
(c) disk-to-halo mass ratio with $\lambda =0.06$ and $F=0.1$. 
The virialized halo is Hernquist halo with core size $c=4$. 
The angular momentum distribution function is 
$f(b,\ell)=(1+10b)\ell^{4/3}-10b\ell^{22/15}$. 
The different  curves correspond to $b= 0, 0.1, 0.3, 0.6, 1$.
The disk-to-halo mass ratio increases but not significantly when 
approaching the centre.}
\end{figure}

\begin{figure}
\centerline{\psfig{file=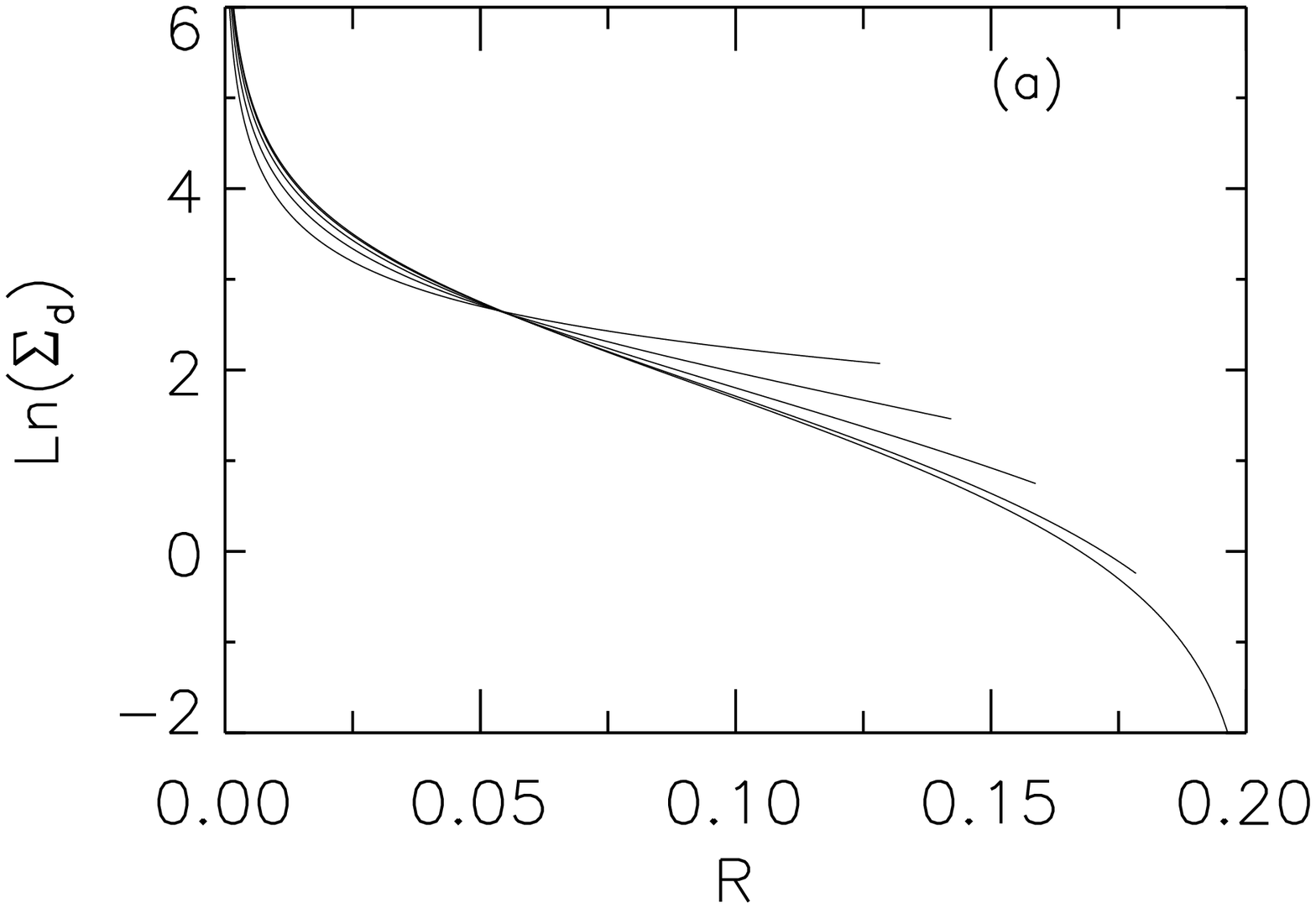,width=2.6in,angle=0}}
\hspace{0.5cm}

\centerline{\psfig{file=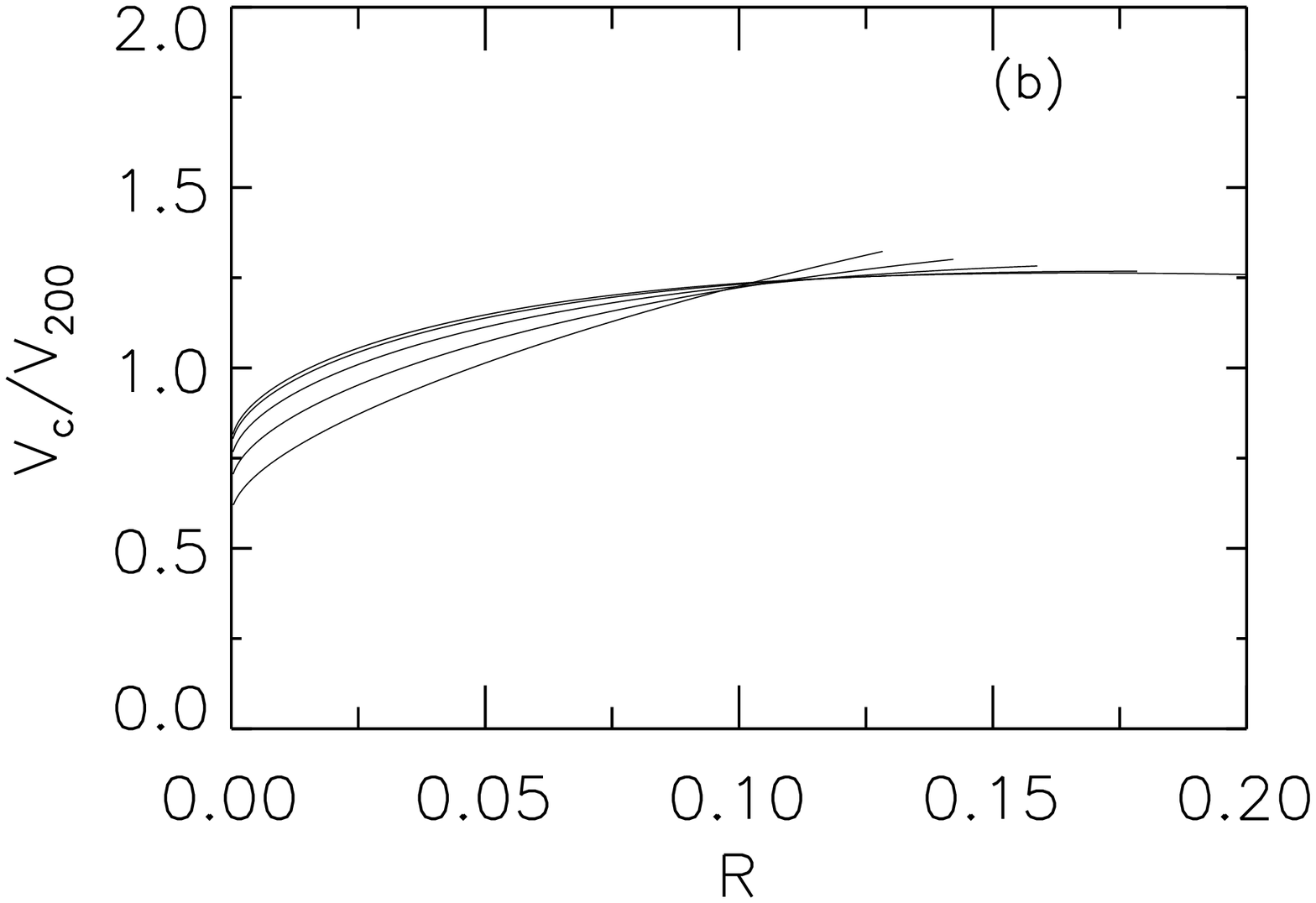,width=2.6in,angle=0}}
\hspace{0.5cm}

\centerline{\psfig{file=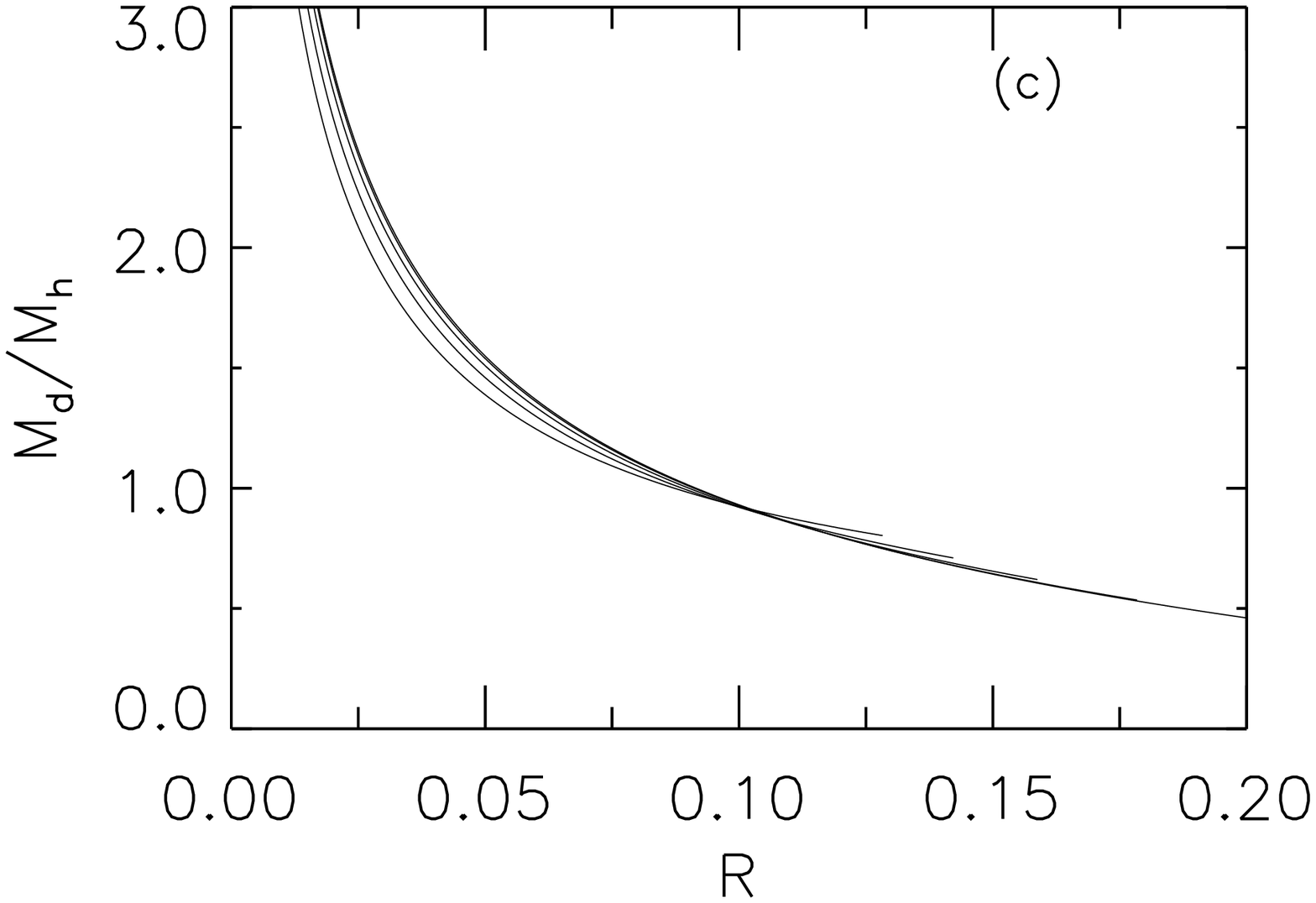,width=2.6in,angle=0}}
\hspace{0.5cm}
\caption{The model corresponding to point D in Figure 1. (a) the surface 
density, (b) circular velocity and 
(c) disk-to-halo mass ratio with $\lambda =0.06$ and $F=0.1$. 
The virialized halo is non-singular isothermal halo with constant 
density core size $c=4$. 
The angular momentum distribution function is 
$f(b,\ell)= (1+b) \ell -b \ell^2 $. 
The different curves correspond to $b= 0, 1/4, 1/2, 3/4, 1$.  
The disk-to-halo mass ratio increases significantly when approaching 
the centre.}
\end{figure}

\begin{figure}
\centerline{\psfig{file=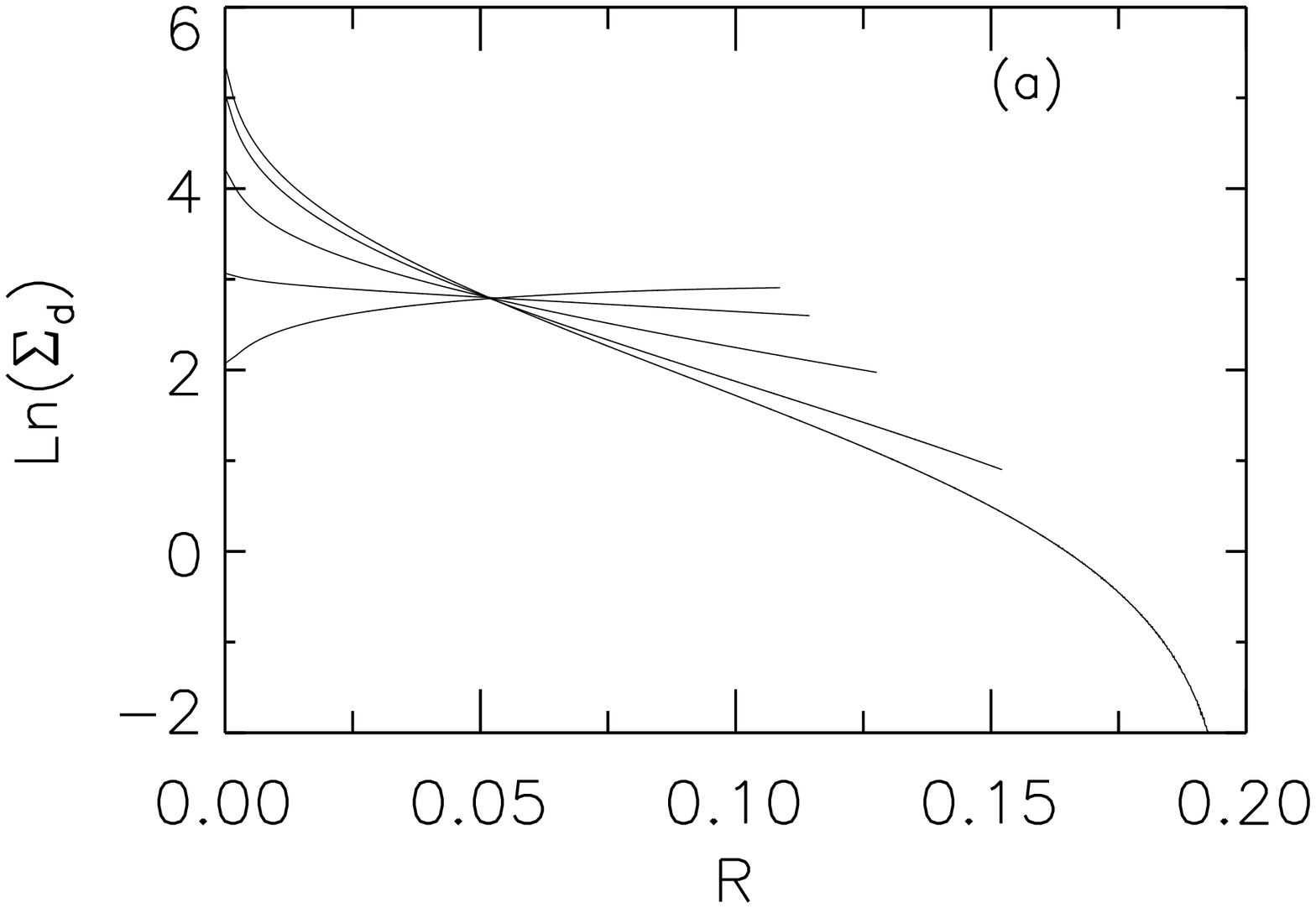,width=2.6in,angle=0}}
\hspace{0.5cm}

\centerline{\psfig{file=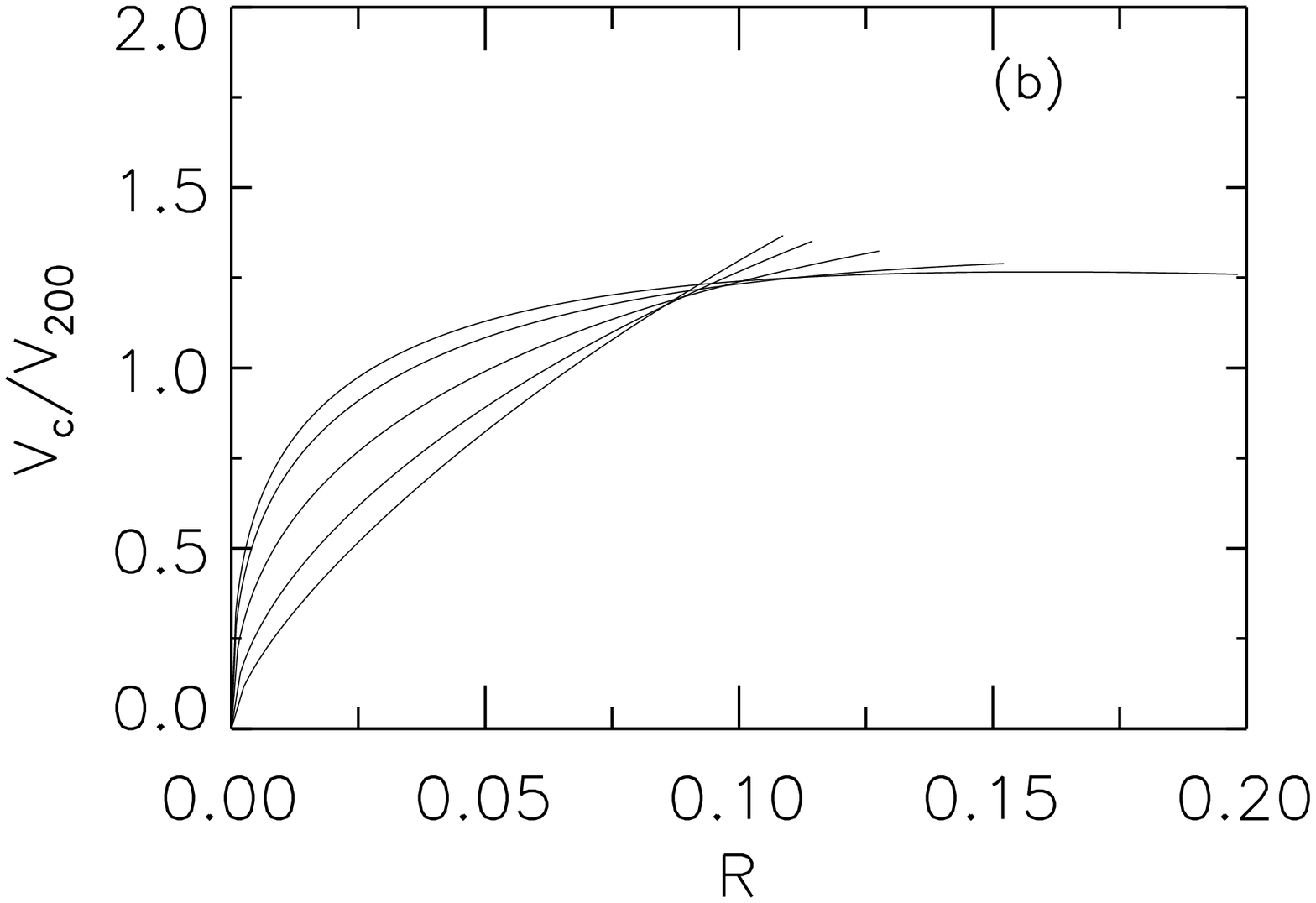,width=2.6in,angle=0}}
\hspace{0.5cm}

\centerline{\psfig{file=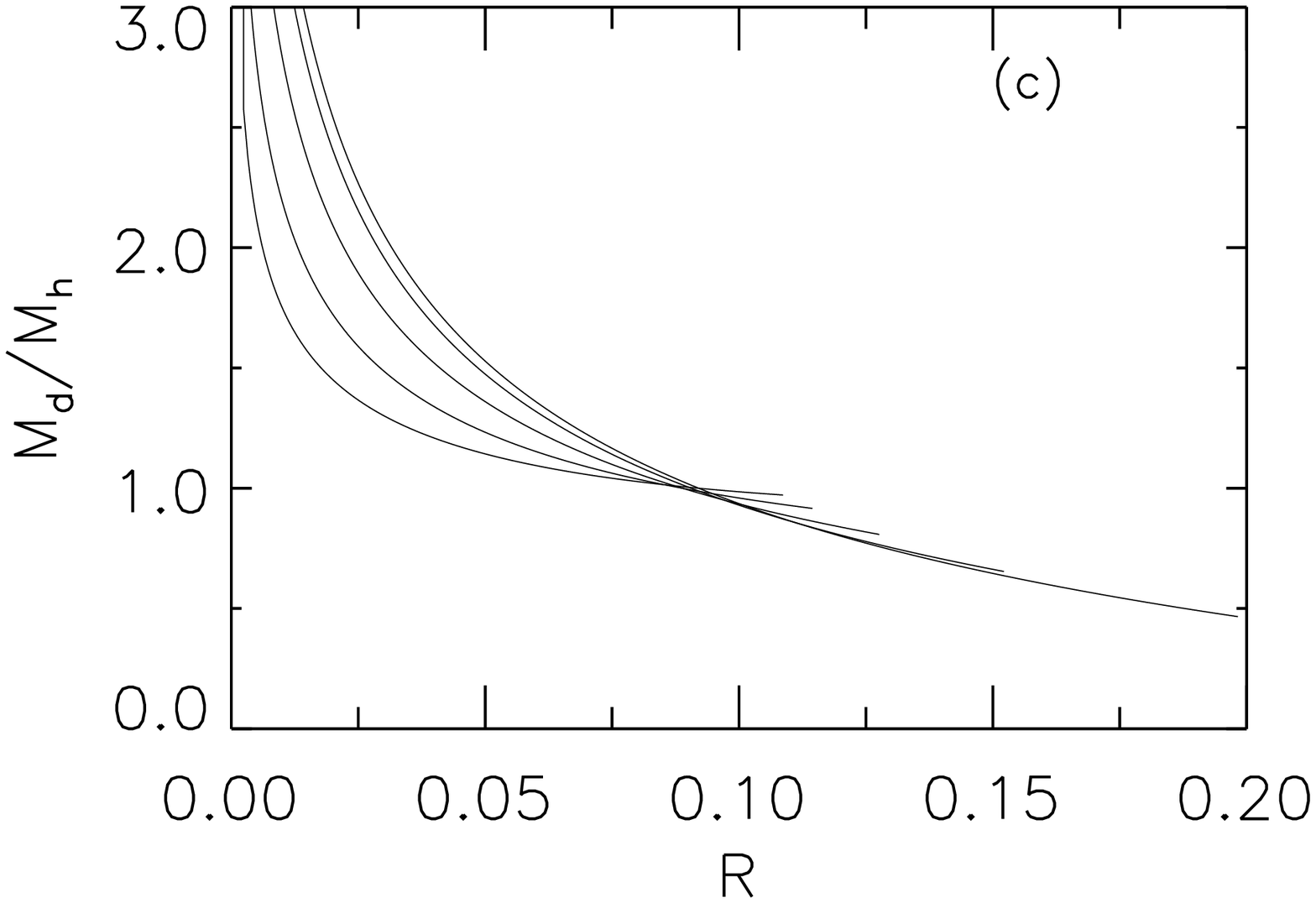,width=2.6in,angle=0}}
\hspace{0.5cm}
\caption{The model corresponding to point C in Figure 1. (a) the surface 
density, (b) circular velocity and 
(c) disk-to-halo mass ratio with $\lambda =0.06$ and $F=0.1$. 
The virialized halo is non-singular isothermal halo halo with 
core size $c=4$. 
The angular momentum distribution function is 
$f(b,\ell)=(1+10b)\ell^{4/3}-10b\ell^{22/15}$. 
The different  curves correspond to $b= 0, 0.1, 0.3, 0.6, 1$.
The disk-to-halo mass ratio increases significantly when approaching 
the centre.}
\end{figure}

The range of viable models represented by the shaded region in Figure 1 can 
be investigated by the appropriate virialized halo profile 
$g(R)$ and angular momentum 
distribution function $f(b,\ell)$ corresponding to the points E,B,D and C. 
The distribution of surface density, circular velocity and disk-to-halo mass 
ratio for these models, varying parameter $b$, are shown in Figures 4--7, 
by choosing fixed $\lambda =0.06$, $F=0.1$ (and 
halo core size $c=4$ if the halo has a core radius).

Model E: The results of a model corresponding to point E in Figure 1 
are shown in Figure 4; this has 
a Hernquist halo profile $g(R) = \frac{(1+c)^2 R^2}{(1+cR)^2}$ and 
$f(b,\ell)= (1+b) \ell -b \ell^2 $. The different curves correspond 
to $b= 0, 1/4, 1/2, 3/4, 1$.

Model B: The results of a model corresponding to point B in Figure 1 
are shown in Figure 5; this has a  
Hernquist halo profile $g(R) = \frac{(1+c)^2 R^2}{(1+cR)^2}$ and 
$f(b,\ell)=(1+10b)\ell^{4/3}-10b\ell^{22/15}$. This $f(b,\ell)$ is shown in 
Figure 2b.
For $ \ell \ll 1$, $f(b=0,\ell) \sim \ell^{4/3}$; the 
$22/15$ index in the second term is determined by the requirement
that $f(b,\ell)$ should be a monotonic increasing function of $\ell$ 
for all values of 
$0 \leq b \leq 1$. The different curves correspond to $b= 0, 0.1, 0.3, 0.6, 1$.

Model D: The results of a model corresponding to point D in Figure 1 
are shown in Figure 6; this has a 
non-singular isothermal halo with a constant density core,  
$g(R) =  \frac{cR-arctan(cR)}{c-arctan(c)} $,  and 
$f(b,\ell)= (1+b) \ell -b \ell^2 $. Again the different curves correspond 
to $b= 0, 1/4, 1/2, 3/4, 1$.

Model C: The results of a model corresponding to point C in Figure 1 
are shown in Figure 7; this has a  
non-singular isothermal halo with constant density core,  
$g(R) =  \frac{cR-arctan(cR)}{c-arctan(c)} $,  and 
$f(b,\ell)=(1+10b)\ell^{4/3}-10b\ell^{22/15}$. 
Again the different  curves correspond to $b= 0, 0.1, 0.3, 0.6, 1$.

As can be seen from the figures, the different halo profiles and angular 
momentum distributions produce disks with a variety of surface density 
profiles and rotation curves. 
As in Figure 3b, the circular velocities 
for points beyond the cut-off radius of a given model may be obtained by 
forming the envelope of the values for the cut-off radius for larger values 
of $b$. 
Thus if the disk is very compact, from equations (27)-(28) or
equation (38), we find that the circular velocity beyond the edge of the disk 
tends to decreases 
with radius.  The location of the edge of the disk depends on both  
$\lambda$ and $F$ (in addition to $b$).  Thus we have shown that 
rotation curves should show an imperfect disk--halo `conspiracy' if the disk 
is too compact or 
too massive. This is consistent with observations 
(Casertano \& van Gorkom 1991).

Further, these figures demonstrate that with increasing $b$, the 
inner rotation curves become flat, and the 
transition between disk-dominated and halo-dominated sections of the 
rotation curve  becomes more and more smooth with increasing 
$b$. As discussed above, and illustrated in Figure 2, a higher value of
$b$ corresponds to a flatter specific angular momentum distribution 
function, and increasing $b$ mimics the effects of viscous evolution
in transporting angular momentum. This indicates that viscous evolution 
can help the creation of an apparent disk-halo `conspiracy'.

As can be seen from equations (21) - (26), for given $c_0$ (the 
compactness parameter defined in equation (20)), or 
$F/\lambda$ ratio, within a given model of $f(\ell,b)$ 
and virialized dark halo profile 
$g(R)$, the normalized properties of the disks formed for 
$b=0$ are very similar.  In particular the disk 
surface density profile, rotation curve and disk-to-halo mass 
ratio profile scale similarly with $\ell$. 
For example, in the case of the singular isothermal sphere (model F), 
equations (34) - (37) show that  for $b=0$ the 
normalized disks 
are identical for given $F/\lambda$ ratio. Thus $F/\lambda$ 
must be an important factor in distinguishing one disk from another. 
The overall normalization of the surface density is $\propto F/\lambda^2$, 
so that 
again $F/\lambda$ and $\lambda$ enter separately and are both 
important.

Note that here we are not insisting that the surface density profile 
of the gas disk 
so formed be exponential, unlike previous work (Mo, Mao \& White 1998).  
We shall however appeal to viscous 
evolution tied to star formation to provide a stellar exponential disk. We 
now turn to this. 

\section{The Viscous Evolution and Star Formation}

In this paper we aim to link viscous evolution within 
disks and the Hubble sequence of disk galaxies.  One of the motivations for 
invoking viscous disks is that if the timescale of angular momentum 
transport via viscosity is similar to that of star formation, a stellar 
exponential disk is naturally produced independent of the initial gaseous disk 
surface density profile (Silk \& Norman 1981; 
Lin \& Pringle 1987; Saio \& Yoshii 1990; Firmani, Hernandez \& Gallagher 1996).   
Angular momentum transport and associated radial gas flows (both inwards and 
outwards) can also, as shown above, provide a tight `conspiracy' between 
disk and halo rotation curves, and, as demonstrated below, provide a higher 
phase space density in bulges as compared to disks.

The star formation rate per unit area in a disk  
can be represented by a modified 
Schmidt law involving the dynamical time and the gas density (Wyse 1986; 
Wyse \& Silk 1989).  We shall use the form of the global star formation rate 
per unit area, $\Sigma_{\psi}$, of Kennicutt (1998), based on his 
observations of the inner regions of nearby large disk galaxies: 
\beq
\Sigma_{\psi} =\alpha \Sigma_{gas}\Omega_{gas},
\eeq
where $\Sigma_{gas}$ is total gas surface density,  $\Omega_{gas}$ is 
the dynamical time at the edge of the gas disk, and the normalization 
constant, related to the efficiency of star formation, has the value 
$\alpha=0.017$ (Kennicutt 1998). Note that observations of the star-formation 
rates in the outer regions of disk galaxies suggest that it is actually 
{\it volume\/} density that should enter the Schmidt law, rather than 
surface density, and since many (if not all) gas disks flare in their 
outer regions, equation (40) will over estimate the star formation rate 
(Ferguson {\it et al.} 1998). This is beyond the scope of the present model, 
but should be borne in mind and will be incorporated in our future work.

For our models here, the edge of the initial gas disk is where $\ell=1$, 
$R=R_c$ and thus $\Omega_{gas} = \Omega_c$. 
For halo formation redshift $z_f$, 
we can obtain the relationship between global star formation timescale 
and the galaxy initial conditions 
in the general form using equation (1), (25) and (39):
\begin{eqnarray}
t_*^{-1}&=&\alpha \Omega_c 
= \alpha V_c(r_c)/r_c  \nonumber\\
&=& 10 \alpha H(z_f) \xi (1-F)^2(\xi m_{hc}+c_0)^2,
\end{eqnarray}
where  again $m_{hc}$ is the solution of $1= \xi^2 m_{hc} g^{-1}(m_{hc})$, 
and is the fraction of the dark halo mass that is contained within the 
cut-off radius of the disk.  The parameters $\xi$ and $c_0$ are defined in 
equations (13) and (20); $F$ is the baryonic mass fraction in the initial density 
perturbation. 

In the case of the singular isothermal halo, this relation has a simple 
form: 
\beq
t_*^{-1}= 10 \alpha H(z_f) \xi (1-F)^2(1+c_0)^2.
\eeq
The gas consumption timescale will be longer than the characteristic 
star formation timescale due to the gas returned by stars during their 
evolution and death.  For a standard stellar Initial Mass Function, 
$t_g \sim 2.5t_*$ (e.g. Kennicutt {\it et al.} 1994).

This modified Schmidt law is based on observations of the inner regions of
nearby large disk galaxies, and simple theoretical principles.
Assuming it holds at all epochs allows one to estimate the  properties 
of present-day disks from 
the initial conditions of earlier sections in this paper. 
Let us assume that the dark halo is fully virialized at redshift $z_f$. 
In keeping with the spirit of hierarchical clustering, let us allow for 
some star formation that could have taken place in the disk, from an earlier 
redshift $z_i$, and that the total mass of the system could increase until 
$z_f$ (although to maintain the thin disk, this accretion and merging must 
be only of low mass, low density systems).  
So for any time $t$ or redshift $z$ between $z_i$ and $z_f$,
\begin{eqnarray}
\frac{d M_g}{d t} &=& F \frac{d M_{tot}}{d t} - \frac{d M_*}{d t}, \\
\frac{d M_*}{d t} &=&  \frac{M_g}{t_g(z)},
\end{eqnarray}
where $M_{tot}$ is defined in equation (1) and $M_*$ is mass locked up into 
stars.
In an Einstein-de-Sitter Universe, $H(z)= H_0 (1+z)^{3/2}$ and
$H(z)t= 2/3$.
We have
\beq
\frac{d M_g}{dt} = A -\frac{B M_g}{ t},
\eeq
where $A=\frac{3F V^3_{200}}{20 G} $ and 
$B=\frac{8}{3} \alpha \xi(1-F)^2(1+c_0)^2 $, and $\xi$ and $c_0$ are 
expressed in terms of $c_f$, $\lambda$ and $F$ through equations (13) and 
(20) in section 2.1. Hence, from equation (1), 
$M_{tot} \propto \frac{A}{F H(z)}$. 

Identifying the dark halo to have fixed $V_{200}$, independent of redshift, 
leads to $A$ also being a constant, and thus the total mass grows as 
$M_{tot} \propto t$.  This differs from the standard solution of infall onto 
a point, $M_{tot} \propto t^{2/3}$ (Gunn \& Gott 1972).

The solution to the above equation is then 
\beq
M_g = \frac{A}{1+B} t \left[ 1+ B\left( \frac{t}{t_i} \right)^{-(1+B)} \right], 
\eeq
where $t_i$ corresponds to the redshift $z_i$ of the onset of star 
formation. Thus at the halo formation redshift $z_f$, the disk gas fraction is 
\beq
f_g(z_f) = \frac{1+B \left( \frac{1+z_f}{1+z_i} \right)^{3(1+B)/2}}{1+B},
\eeq
Since $B$ depends on $c_f$, $\lambda$ and $F$, i.e. 
$B \propto \frac{c_f}{\lambda} (1-F +\frac{c_f}{\sqrt{2}} \frac{F}{\lambda})^2$, 
the value of the constant $B$ may be evaluated 
for  $\lambda =0.06$ and various reasonable values of $F$ and $c_f$ as: 
for $F=0.1$, $c_f= 1/3$, $B=0.30$;
for $F=0.05$, $c_f= 1/3$, $B=0.23$;
for $F=0.1$, $c_f= 0.5$, $B=0.59$;
and for $F=0.05$, $c_f= 0.5$, $B=0.41$.
For fixed $\lambda$ and $F$, small values of $B$ correspond to small values 
of $c_f$, and hence flatter specific angular momentum distributions.

Thus for $1+z_f  \lta 2(1+ z_i)$, $f_g(z_f) \approx 1/(1+B)$, typically $\gta 2/3$.

We assume that after $z_f$, there is no further infall, and that  
the gas in the disk will be consumed with characteristic timescale $t_g(z_f)$.
Thus the gas fraction of a typical disk galaxy  at the 
present time is 
\beq
f_g=f_g(z_f) \exp \left[ -\delta t(z_f)/t_g(z_f)\right],
\eeq 
where $\delta t(z_f)$ is the time interval between the halo formation redshift $z_f$ 
and present time $z=0$. 

Assuming an Einstein-de-Sitter universe, 
we have $\delta t(z_f)= t_0(1-H_0/H(z_f))$ and $t_0 H_0 = 2/3$. 
Thus a typical value for the present gas fraction of disk galaxies is:
\beq
\ln \left( \frac{f_g(z_f)}{f_g} \right) = B \left[ (1+z_f)^{3/2}-1 \right]. 
\eeq
With the approximation $f_g(z_f) = 1/(1+B)$, we have
\beq
(1+z_f)^{3/2} = 1+ \frac{-\ln f_g- \ln (1+B)}{B}.
\eeq

For the Milky Way Galaxy, if we adopt $5 \times 10^9 M_{\odot}$ for the atomic 
$HI$ gas, and $1.3 \times 10^9 M_{\odot}$ for the molecular $H_2$ gas 
(Blitz 1996; Dame 1993), then with an estimate of 
$4 - 6 \times 10^{10} M_{\odot}$ for the total baryonic mass of the Milky Way 
depending on stellar exponential scale length (Dehnen \& Binney 1998), 
we obtain gas fraction, the total gas mass includes $24 \%$ helium mass, 
$f_g \sim 15\%$ or even higher depending on the mass model of 
the Galaxy. As mentioned earlier, for fixed $\lambda$ and $F$, 
small values of $B$ correspond to small values 
of $c_f$, and hence flatter specific angular momentum distributions.
For small values of the parameter $B$, say $B \sim 0.3$, we obtain 
$z_f \sim 2.4$, while for large values of $B$,  $B \sim 0.6$, 
we obtain $z_f \sim 1.2$. 
The larger value of $z_f$ is preferred, given what we know of the 
age distribution  of stars in the local thin disk (e.g. Edvardsson \et 
1993). The effect of viscous evolution is equivalent to choosing 
small $c_f$, i.e. 
small values of $B$. So the inclusion of viscous evolution can give relatively 
higher halo formation redshift, which is consistent with the results from the 
constraint on the redshift of formation that we obtained from considerations 
of the size of the disk.

It should be noted that for fixed halo `formation' redshift $z_f$,
the star formation timescale $t_*^{-1} \sim \frac{1}{\lambda} 
(1-F+\frac{1}{2 \sqrt{2}} \frac{F}{\lambda})^2$. 
Again, both 
$\lambda$ and $F/\lambda$ are important. 
As we discussed in the previous section, the structure of the normalized 
disk 
depends strongly on $F/\lambda$ while the overall normalization depends strongly
on $\lambda$ for fixed $F/\lambda$. 
As we show later, 
the bulge-to-disk ratio also 
depends on the importance of $\lambda$ and $F/\lambda$.
Thus many aspects of the Hubble sequence of disk galaxies  -- 
star formation timescale,   
disk gas fraction, bulge-to-disk ratio -- depend on both $\lambda$ and 
$F/\lambda$.

The star formation timescale derived above is independent of $V_{200}$, 
which at first sight is surprising, given that the 
Hubble sequence of disk galaxies has been interpreted as a sequence of 
star formation timescales, relative to collapse times (Sandage 1986), and 
observations show that the Hubble type of a disk galaxies is broadly 
correlated with the disk luminosity (Lake \& Carlberg 1988; de Jong 1995).
However, $V_{200}$ is not an easily-observed quantity. 

The present-day luminosity of a disk in our model can be written as
\beq L = \frac{F M_{tot}(1-f_g)}{\gamma_*}=\frac{F[1-f_g(B,z_f)]
V_{200}^3}{10GH(z_f) \gamma_*}, \eeq with $\gamma_*$ the current value
of the mass-to-light ratio.  Estimation of the predicted Tully-Fisher relation
depends on what model parameter we use for the width of the
HI line.  Obviously if we identify this with $V_{200}$ we have a
reasonable relationship provided that the coefficient $\gamma_{tf}$ is constant,
i.e. $F[1-f_g(B,z_f)]/H(z_f) = \gamma_{TF}$.  This then requires that there
should be a correlation between $F$ and $z_f$, in the sense that for
large $F$, $z_f$ is large.  Then we can see from equation (42) that
this leads to the star formation timescale being small, 
implying more efficient viscous
evolution, and leading to larger B/T ratio. Large $z_f$ may be
correlated with small $V_{200}$ in the context of
hierarchical-clustering cosmologies, and in that case, a short 
star formation time and large B/T
ratio should be correlated with low luminosity, which is not
consistent with observation.  However, an interpretation that is compatible with observations is that high n-sigma fluctuations for
fixed $V_{200}$ can form high luminosity disks with large B/T ratio.

The Tully-Fisher relationship is not
a simple relation between luminosity and $V_{200}$, but depends on
where the circular velocity $V_c(R)$ is measured (Courteau 1997).  The
relationship between $V_{200}$ and $V_c(R)$ obviously depends on the
details of the halo density profile and the angular momentum
distribution. From the rotation curves in fig. 3 to fig. 7, we can see
that it is appropriate to choose for our estimate of $V_c$ the
circular velocity at the cutoff radius of disk, adopted as 3 three
scale lengths. From equation (35), we have $V_c=V_{200}(1-F)(1+c_0)$,
where $c_0=c_f F/\sqrt{2}\lambda (1-F)$, as defined in equation (20),
is the compactness of disk. Now the predicted Tully-Fisher relation is
\beq L = \frac{F[1-f_g(B,z_f)] V_c^3}{10GH(z_f)
\gamma_*(1+c_0)^3(1-F)^3}. \eeq Requiring that the coefficient $\gamma_{TF}$ 
in the Tully-Fisher relation be constant, i.e. $\gamma_{TF} \equiv F[1-f_g(B,z_f)]/H(z_f)(1+c_0)^3(1-F)^3$, then gives, from
equation (42), $t_*^{-1} \propto c_0/(1+c_0)$. So for small $c_0$, the
star formation timescale is large, which can cause less efficient
viscous evolution. So less efficient viscous evolution and small
$F/\lambda$ will lead to small B/T ratio. Also small $c_0$ is
correlated with large $z_f$ from the constancy of the coefficient in
the Tully-Fisher relation. Similarly large $z_f$ may be correlated
with small $V_{200}$ in the context of hierarchical clustering
cosmology. So small $c_0$ and small $V_{200}$ will lead to small $V_c$
and lower disk luminosity. Thus this version of the predicted Tully-Fisher relation appears fully compatible with the observations. 
However, one should bear in mind that we have adopted a fixed constant of
proportionality $\alpha$ in the star formation law, and this may well
vary with global potential well depth (White \& Frenk 1991) or local
potential well depth (Silk \& Wyse 1993).

A further test of the model is the relation between disk scale and
circular velocity, and its variation with redshift; observations
indicate that $R_d/V_c$ is smaller at high redshift, $z \sim 1$ (Vogt
{\it et al.} 1996; Simard {\it et al.} 1999).  In our model there is
little change in total mass between redshifts of unity and the
present, and so this evolution of disk size cannot be due to the halo
mass growth, as had been proposed by Mao, Mo \& White (1998). Instead
in our model this is due to the different and changing scale lengths of
gas and stars. In our model, due to early star formation, the gas
fraction at $z_f$ is about $f_g(z_f) \sim 1/(1+B)$, typically
$2/3$. It is natural  that the gas component of the disk will 
have a larger scale length than the stellar component. The scale
length of the gas component can increase with time due to viscous
evolution while the scale length of stellar component can also
increase with time due to the non-linear local star formation law
(Saio \& Yoshi 1990). So the stellar scale length at high redshift,
when the gas fraction is large, should be much smaller than the
stellar scale length at present time, when gas fraction is lower. 
The study of detailed evolution of gas and stellar component is 
beyond the scope of this paper; but the prediction of stellar size 
evolution of our model is qualitatively consistent with the observations.
 
The measured distribution of $R_d/V_c$ for the local disk galaxy sample is approximately peaked at $R_d/V_c \sim -1.5$ ($R_d$ in kpc, $V_c$ in kms$^{-1}$) with a spread from -2 to -1 (Mao, Mo \& White 1998; Courteau 1996). 
Our estimation of the disk size or scale length in Section 2 is valid for
galaxies at the present time, when the gas fraction is small. Then
assuming the disk cut off radius is three scale lengths, 
from equation (41), $R_d/V_c =R_c/3V_c=\alpha t_*/3$, 
and the present $R_d/V_c $ is an indication
of the galactic global star formation timescale; 
further, from the constancy of $\gamma_{TF}$, the coefficient of the Tully-Fisher relation in equation (52), we have 
\beq R_d/V_c =\alpha t_*/3=\frac{G}{3} \gamma_{TF} \gamma_*\frac{1+c_0}{c_0(1-f_g)}. \eeq 
Thus this ratio is also an indication of the disk compactness. 
Adopting the B-band mass-to-light ratio of our local disk $\gamma_* \sim 2.5 M_{\odot}/L_{\odot}$, and using the luminosity of our galaxy 
$L_B \sim 3 \times 10^{10} L_{\odot}$ and $V_c \sim 220$ kms$^{-1}$ to estimate 
$\gamma_{TF}=L_B/V_c^3$, we can obtain that  $\log(R_d/V_c) \simeq -2.0+\log(\frac{1+c_0}{c_0(1-f_g)})$. Obviously $\log(R_d/V_c) \sim -2$ roughly corresponds to the predicted  lower limit of the local sample, which is consistent with observations. 
The distribution of $F$ from 0.05 to 0.2 and the distribution of $\lambda$ from 0.03 to 0.12 will cause the value of the compactness parameter $c_0$ to spread from 0.1 to 2 approximately. Adopting typical values $F=0.1$, $\lambda=0.06$, $c_f=1/3$, the peak will be located at $\log(R_d/V_c) = -1.5$ with $c_0=0.44$, which is again consistent with observations. The spread of $R_d/V_c$ is simply caused by the spread in the  value of compactness parameter $c_0$.

\section{The Formation of Bulges and the Hubble Sequence}

Galaxies classification schemes based on morphology are the basic first step 
in understanding how 
galaxies form and evolve (van den Bergh 1998). 
The bulge-to-disk luminosity ratio is 
one of the three basic classification criterion for the 
Hubble sequence (Sandage 1961). 
However, the relation between bulge-to-disk ratio and Hubble type 
has a fair amount of scatter, some of which must be related to the 
difficulty of a decomposition of the light profile into bulge and disk, 
and the bulge-to-disk 
ratio is dependent on the band-pass used to define the luminosity 
(de Jong 1996). 
The current observational data show that bulges can be 
diverse and heterogeneous (Wyse, Gilmore \& Franx 1997). 
Some share properties of disks and some are more similar to ellipticals. 
Models of bulge formation can be classified into 
several categories: the bulge is formed from early collapse of low angular 
momentum gas, with short cooling time and efficient star formation 
(Eggen {\it et al.} 1962; Larson 1976 -- who invoked viscosity to
transport angular momentum away from the proto-bulge; van den Bosch 1998); 
the bulge is formed from merging of disk galaxies (Toomre \& Toomre 1972; 
Kauffmann, White \& Guiderdoni 1993); the bulge is formed 
from the disk by secular evolution process 
after bar instability (Combes {\it et al.} 1990; 
Norman, Sellwood \& Hasan 1996; Sellwood \& Moore 1999; 
Avila-Reese \& Firmani 1999); the bulge is formed from early dynamical 
evolution of massive clumps formed in disk (Noguchi 1999).
However it is speculated that large bulges, 
which tend to have de Vaucouleur's law 
surface brightness profiles,  share formation mechanisms with ellipticals, 
while smaller bulges, which tend to be better fit by an exponential profile, 
are formed from their disks through bar dissolution. 
It should be noted that the significantly 
higher phase space density of bulges when compared to disks suggests that 
gaseous inflow should play a part in the instability (Wyse 1998), 
just as invoked in 
earlier sections of this paper, for other reasons.

Here we only consider this latter case, the  formation of small bulges from
the disk. Early studies of disk instabilities showed that Toomre's local
stability criterion $Q > 1$ is also sufficient for global stability to
axisymmetric modes (Hohl 1971; Binney \& Tremaine 1987).  It is known that
the bar instability requires a similar condition (Hockney \& Hohl 1969; Ostriker
\& Peebles 1973). Efstathiou, Lake \& Negroponte (1982) used N-body
techniques to study the global bar instability of a pure exponential disk
embedded in a dark halo  and proposed a simple
instability criterion for the stellar disk, based on the 
disk-to-halo mass ratio.
However, it has been argued recently that there is no such simple 
criterion for bar instability (Christodoulou, Shlosman \& Tohline 1995; 
Sellwood \& Moore 1999).  Further, recent N-body simulations 
(Sellwood \& Moore 1999; Norman, Sellwood \& Hasan 1996) 
show that every massive disk form a bar during the  early stages of 
evolution, but later the bar is 
destroyed by the formation of a dense central object, once the mass of that 
central concentration reaches  several percent of the total disk mass. 
This can be understood in terms of 
the linear mode analysis work and 
nonlinear processes of swing amplifier and feed back loops
(Goldreich \& Lynden-Bell 1965; Julian \& Toomre 1966). Toomre (1981)
argued that the bar-instability can be inhibited in two ways: one is that a 
large 
dark halo mass fraction can reduce the gain of the swing amplifier, while 
the other is that 
feedback through the centre can be shut off by a inner Lindblad 
resonance (ILR). The dense central object can destroy the bar via the second 
of these mechanisms. 
However, most of these studies assume that 
the dark halo has a constant density core. On the contrary, 
we have shown that dark halo profile cannot have a constant density core after 
adiabatic infall, if one starts from physical initial conditions. 
It will be interesting to study  bar formation under 
different disk-halo profiles in addition to the well-studied harmonic core.
 
We will for simplicity adopt the simple criterion of Efstathiou, Lake \& 
Negroponte (1982), interpreted to determine the size of a bar-unstable 
region, with the radial extent of the bar, $r_b$ defined by 
\beq
M_d(r_b)/M_h(r_b) \geq \beta,
\eeq
with the value of the parameter $\beta$ chosen to fit observations. 
As we have argued in section 2, 
$F/\lambda$ is the important quantity determining the overall normalization 
of the disk surface density. 
From N-body simulations (Warren {\it et al.} 1992; Barnes \& Efstathiou 1987; 
Cole \& Lacey 1996; 
Steinmetz \& Bartelmann 1995),
the distribution of $\lambda$ can be fit by a log-normal distribution:
\beq
P(\lambda)d\lambda= \frac{1}{\sqrt{2\pi}\sigma_{\lambda}} 
exp \left[ {-\frac{\ln^2(\lambda/\lambda_0)}
{2\sigma_{\lambda}}}\frac{d\lambda}{\lambda} \right],
\eeq
where $\lambda_0=0.06$ and $\sigma_{\lambda}=0.5$ (this result is fairly 
independent of the slope of the power spectrum of density fluctuations). 

What of the possible range of the baryonic fraction $F$? 
Some 
previous studies suggested that $F \sim \lambda$ as an explanation of the  
disk-halo `conspiracy' 
(Fall \& Efstathiou 1980; Jones \& Wyse 1983; 
Ryden \& Gunn 1987; Hernandez \& Gilmore 1998).
Some interpretations of the observed Tully-Fisher relationship 
suggest that indeed $F$ is not invariant (McGaugh \& de Blok 1998).  
Here we shall assume that the distribution of $F$,  similar to $\lambda$,
 is log-normal,  
centreed at $F_0=0.1$ with
$\sigma_{F}=0.05$. So $F$ is mainly within the range $0.05 \sim 0.2$. 

We generated a Monte Carlo sample of disks, with fixed 
halo $V_{200}$ and halo formation
redshift $z_f$, but with values 
of $F$ and $\lambda$ following the above distributions. 
Then for given virialized dark halo profiles $g(R)$ and 
angular momentum distribution $f(b,\ell)$ in Figure 1,
we can calculate the star formation timescale from equation (41).
The parameter $b$ represents the efficiency of viscous evolution; on the 
assumption that 
the viscous timescale is equal to the star formation timescale, we can 
use a simple linear correlation between the value of $b$ and the 
star formation timescale.  Then the value of parameter $b$ can be obtained. 
From equations (21), (22), (23), (26), (50) and (54), one can calculate 
the bulge-to-disk ratio and final disk gas fraction.

Thus we can plot B/T ratio versus disk gas fraction, for the whole Monte 
Carlo 
sample, confining the parameter space of $\lambda$ and $F$ to give 
small bulges with $B/T < 0.5$. 
Larger values of B/T would correspond to such an unstable disk that the
exercise is invalid.  Disks that are too stable, having low $F/\lambda$ or
large $\lambda$, will evolve little and probably end up as low surface
brightness systems.

\begin{figure}
\centerline{\psfig{file=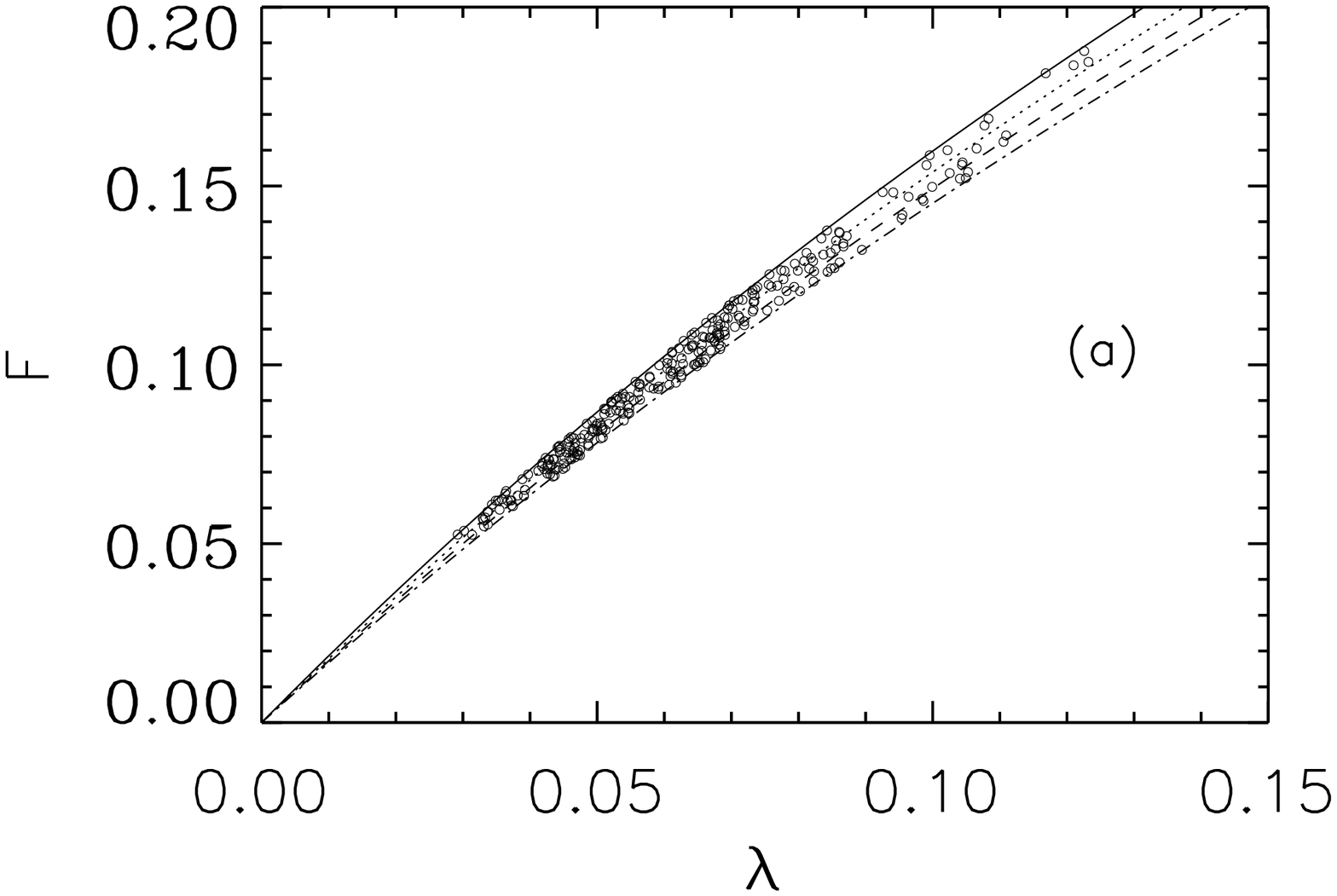,width=2.6in,angle=0}}
\hspace{0.5cm}

\centerline{\psfig{file=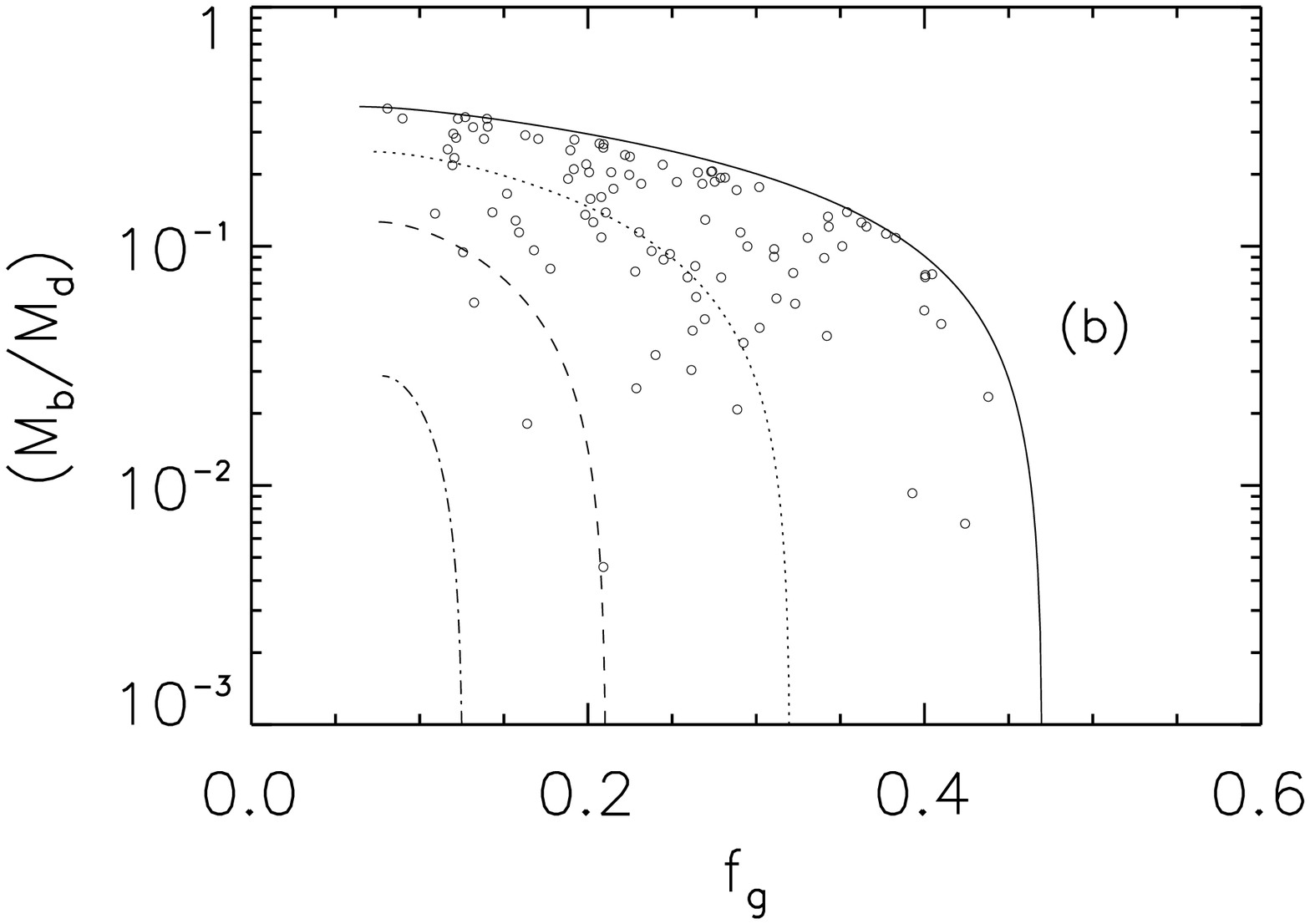,width=2.6in,angle=0}}
\hspace{0.5cm}

\caption{The model corresponding to point F in Figure 1. 
(a) The sample of disk galaxies 
with different $\lambda$ and $F$.
(b) The relation between B/T ratio and gas fraction for these sample
galaxies at current age. The different curves in this plot correspond to
different values of $(1-F)\lambda/F \approx \lambda/F$, which sets the overall trend.}
\end{figure}

\begin{figure}
\centerline{\psfig{file=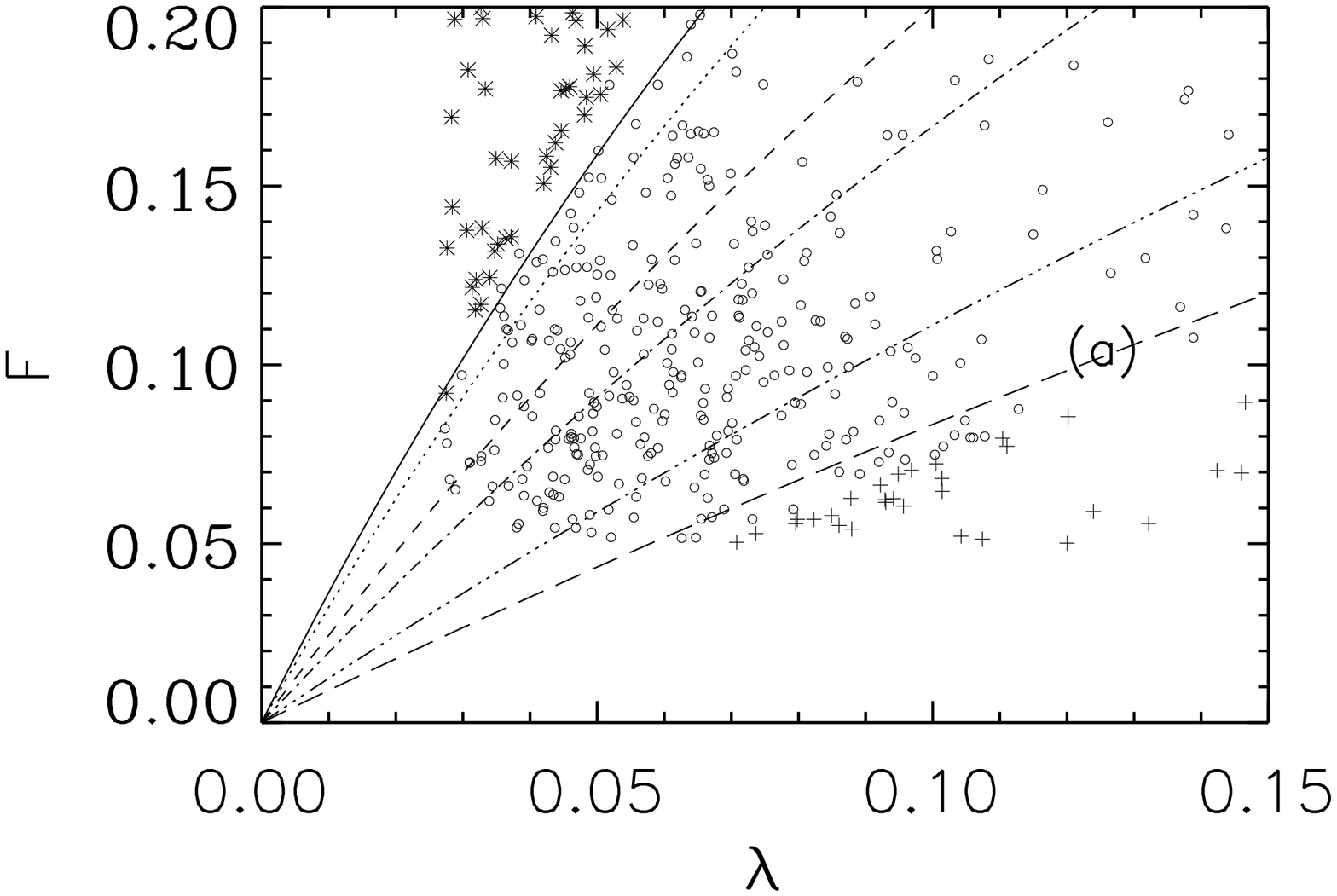,width=2.6in,angle=0}}
\hspace{0.5cm}

\centerline{\psfig{file=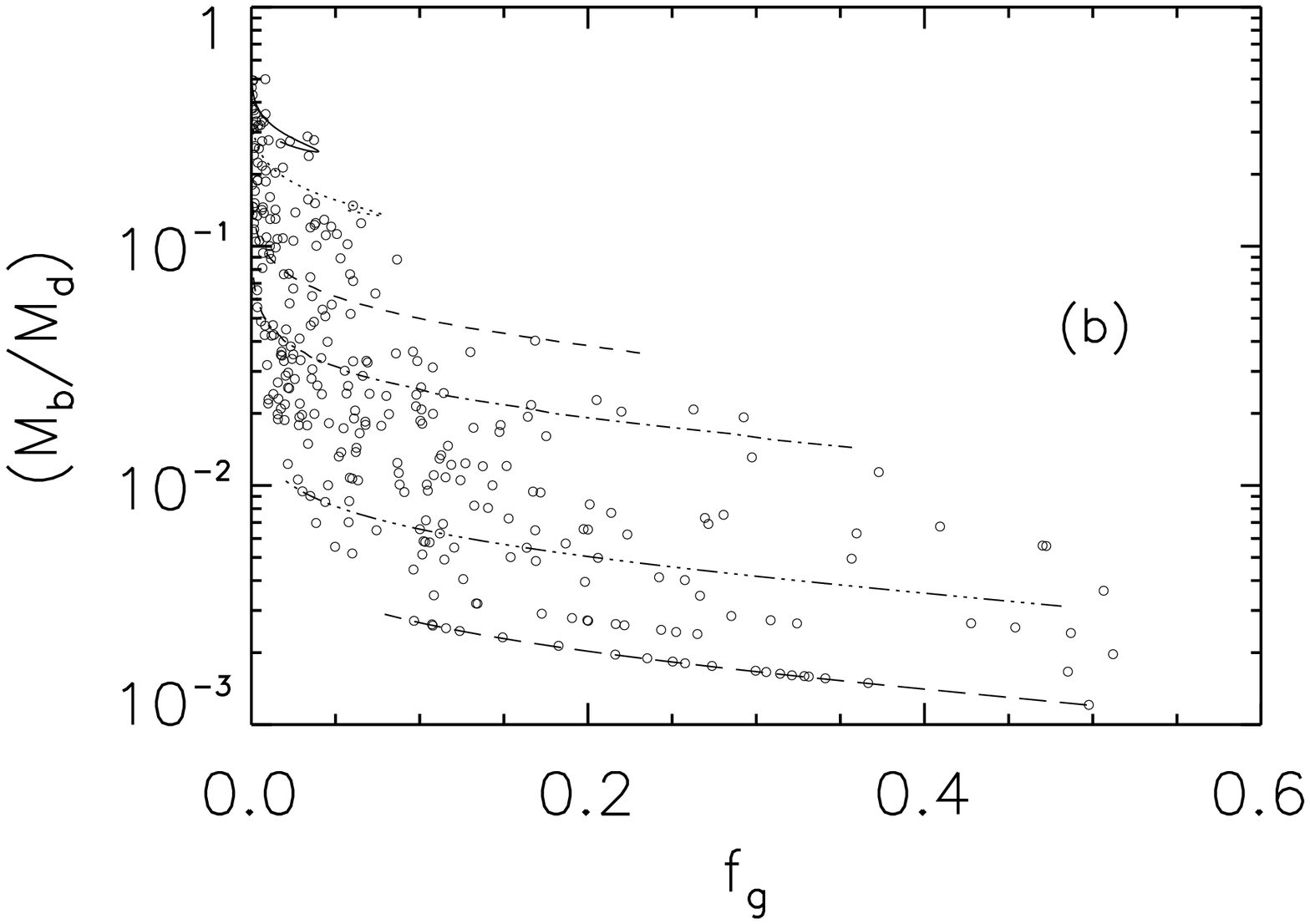,width=2.6in,angle=0}}
\hspace{0.5cm}

\caption{The model corresponding to point E in Figure 1. 
(a) The sample of disk galaxies 
with different $\lambda$ and $F$. Overly-unstable disks are denoted by asterisk symbols. Bulgeless disks or low surface-brightness disks are denoted by cross symbols.
(b) The relation between B/T ratio and gas fraction for these sample
galaxies at current age. The different curves in this plot correspond to
different values of $(1-F)\lambda/F \approx \lambda/F$, which sets the overall trend.}
\end{figure}

Figure 8a shows  model F, representing the singular isothermal halo. 
Assuming the typical values $\lambda=0.06$ and $F=0.1$, then 
the choice of $\beta \sim 0.8$ is required to  allow the existence of 
bulges.
This is just the value of $\beta$ used in the 
global bar instability criterion (Efstathiou, Lake \& Negroponte 1982).  
For this model the disk-to-halo mass ratio varies little with radius (as 
shown in Figure 3a), so the B/T ratio is strongly 
dependent on $\lambda/F$. Only a small range of values of $\lambda/F$ is 
allowed, so as to 
not over-produce either bulge-less disks or completely unstable disks.

The relation between B/T ratio and gas fraction for this model is given in 
Figure 8b; there a general trend in the observed sense, 
with large scatter.  The different curves in this plot correspond to
different values of $\lambda/F$, which sets the overall trend. 

Figure 9 (a,b) shows the equivalent plots of model E, 
representing a Hernquist profile halo model, which is probably a 
more realistic case.  Here the  
the disk-to-halo mass ratio varies strongly with radius when approaching the 
centre, and the allowed parameter space for values of 
$\lambda$ and of $F$ that allowing the formation of
bulges can be large.   Overly-unstable disks are denoted by asterisk symbols 
in Figure 9a, and 
low surface-brightness disks are denoted by cross symbols. 
The relation between B/T ratio and
gas fraction is similar to that for the isothermal halo. 

\subsection{Constraints from the Milky Way}

The Milky Way bulge is reasonably well-fit by an exponential profile, with a 
scale-length approximately one-tenth that of the disk (Kent {\it et al.} 1991). The morphology 
of the bulge is consistent with some triaxiality (Blitz \& Spergel 1991; Binney {\it et al.} 1997).  Perhaps the Milky 
Way is a system in which the bulge has formed from the disk, through a bar 
instability? 
Observations show no evidence for a significant young or even 
intermediate-age stellar population in 
the field population of the 
Galactic Bulge (Feltzing \& Gilmore 1999), despite their being ongoing star 
formation in the inner disk. This implies that 
if the bulge were formed from the disk through bar dissolution, 
only one 
such episode is allowed, and this should have happened at high redshift. 
In the context of the present model, the lower star formation rates and 
longer viscosity timescales of later times act to stabilize the system. 
However, it remains to be seen if the observed relative 
frequencies of bars, bulges and central mass concentration is consistent 
with the models of bar dissolution.

\section{Summary}

In the context of hierarchical clustering cosmology, the 
dark halo of a disk galaxy can be formed by quiescently merging small 
sub-halos into the primary dark halo, or by smoothly accreting matter 
into the 
dark halo. We derive the generic solution to the adiabatic infall model of
disk galaxy formation pioneered by many authors 
(Mestel 1963; Fall \& Efstathiou 1980; 
Gunn 1982; Faber 1982; Jones \& Wyse 1983; Ryden \& Gunn 1987; 
Dalcanton, Summer \& Spergel 1997; Mo, Mao \& White 1998; 
Hernandez \& Gilmore 1998). 
Through exploring the allowed parameter space of dark halo profile and angular 
momentum distribution function, we show that the central halo
density profile should be cuspy, with the power law index ranging 
from $-0.75 $ to $-2$, in the central regions where the disk mass dominates.

Using a modified Schmidt law of 
global star formation rate, we derive a simple scaling relationship between
the disk gas fraction and the assembly redshift. 
We explicitly allow a distribution in the values of the baryonic mass 
fraction, $F$, in addition of the 
distribution in values of the spin parameter $\lambda$.
These two are found to play different role in determining the structural 
properties of the final disk, the star formation properties and 
bulge-to-disk ratio. 

We mimic viscous evolution of disks by varying the specific angular momentum 
distribution of the disk, to redistribute angular momentum as a function of 
time.  We derive a consistent picture of the formation of galaxies like the 
Milky Way, with old stars in the disk. 
Under the assumption that the viscous evolution timescale is equal to the star 
formation timescale, we can further combine the $\lambda$ and $F$ with 
the efficiency of angular momentum redistribution caused by viscosity. 
Assuming that small bulges are formed from their disks 
through bar dissolution,
we can use the global bar instability condition to obtain bulge-to-total 
ratio, and explore the dependence on  $F$, $\lambda$ and the viscous 
evolution efficiency.

The inclusion of viscous evolution has the merits of addressing several 
important 
issues:  the conspiracy between disk and halo, the formation of the 
exponential profile 
of stellar disk, the high phase space density of bulges. 
We have presented an analytic treatment, to illustrate these points and 
identify areas of particular need for more work.

\section*{Acknowledgments}
We acknowledge support from NASA, ATP Grant NAG5-3928.
BZ thanks Colin Norman, Jay Gallagher for helpful comments.
RFGW thanks all at the Center for Particle Astrophysics, UC Berkeley, for 
their hospitality during the early stages of this work.

\end{document}